\documentclass[letterpaper,journal]{IEEEtran}
\usepackage{cite}
\usepackage{amsmath,amssymb,amsfonts,mathtools,bm}
\usepackage{booktabs,multirow,array}
\usepackage{enumitem}
\usepackage{graphicx}
\usepackage{url}
\usepackage{tcolorbox}
\usepackage[hidelinks]{hyperref}
\hypersetup{pdftitle={Multimodal Large Language Model-guided Constrained Optimization for RAN Intelligent Control},pdfauthor={Hyeonho Noh}}

\makeatletter
\renewcommand\paragraph[1]{%
    \vspace{.1cm}\noindent\textbf{#1.}
}
\makeatother

\DeclareMathAlphabet{\mathdutchcal}{U}{eus}{b}{n}
\DeclareMathAlphabet{\mathcal}{OMS}{cmsy}{m}{n}
\SetMathAlphabet{\mathcal}{bold}{OMS}{cmsy}{b}{n}
\newcommand{\Problem}[1]{\mathbf{\mathdutchcal{P}}_{#1}}
\newcounter{savedproblemequation}
\newenvironment{problem}[1]
{%
    \setcounter{savedproblemequation}{\value{equation}}%
    \begingroup
    \begin{subequations}%
    \renewcommand{\theparentequation}{\ensuremath{\mathdutchcal{P}_{#1}}}%
}
{%
    \end{subequations}%
    \setcounter{equation}{\value{savedproblemequation}}%
    \endgroup
}

\allowdisplaybreaks[4]

\begin{document}

\title{Multimodal Large Language Model-guided Constrained Optimization for RAN Intelligent Control}
\author{Hyeonho Noh,~\IEEEmembership{Member,~IEEE}
\thanks{Hyeonho Noh is with the Department of Information and Communication Engineering, Hanbat National University, Republic of Korea (e-mail: hhnoh@hanbat.ac.kr). 
}
}

\maketitle

\begin{abstract}
Artificial intelligence (AI)-based radio access network (RAN) controllers are commonly designed for predefined operating scenarios and optimization tasks, limiting their adaptability when network conditions and operator requirements change after deployment. This paper proposes multimodal large language model (MLLM)-guided constrained optimization for RAN intelligent control (MLLM-coRIC), a requirement-adaptive hierarchical Open RAN (O-RAN) framework for joint resource and power allocation. MLLM-coRIC jointly exploits the operator's natural-language specification and radio-frequency (RF)-derived network context as multimodal inputs, enabling a single deployed framework to address different optimization problems without redesigning task-specific algorithms or retraining control policies. At the Non-Real-Time RAN Intelligent Controller (Non-RT RIC), the MLLM performs holistic, longer-timescale reasoning over the operator requirement, predicted network evolution, and measured optimization outcomes to design the numerical control loss. The relative constraint penalties are iteratively refined through a closed-loop process, allowing the optimization criterion to reflect both the intended network behavior and the expected operating context. At the Near-Real-Time RIC (Near-RT RIC), a loss-conditioned hybrid executor translates the synthesized criterion into fast radio actions by combining learned ramp-constrained resource allocation with model-based interference-aware power control. A multi-cell evaluation environment integrating the CARLA urban mobility simulator and the Sionna RT ray-tracing-based wireless propagation simulator is implemented for validation. Results show that MLLM-coRIC consistently outperforms conventional and learning-based baselines across varying user loads, objectives, constraints, and operating thresholds, while adapting to changing optimization problems without per-request retraining.
\end{abstract}


\section{Introduction}
\label{sec:intro}

The evolution toward sixth-generation (6G) networks is driving radio access networks (RANs) from task-specific automation toward artificial intelligence (AI)-native control, where network intelligence is expected to continuously interpret operating conditions and adapt control decisions accordingly \cite{Letaief19_CM,Shi23_COMST,Shafin20_WC,Yang25_VTM}. Open RAN (O-RAN) provides a natural architectural foundation for this transition by combining disaggregated network functions with programmable control interfaces \cite{Polese23_COMST,Bonati21_COMM}. Rather than embedding intelligence into a monolithic RAN stack, control functions can be exposed and coordinated across open interfaces, allowing learning-based network operation to evolve together with changing traffic, service, and operator requirements \cite{doro24_TMC,Tsampazi25_TMC}.

A key feature of O-RAN is the separation of control intelligence across different timescales \cite{polese2023understanding}. The Non-Real-Time RAN Intelligent Controller (Non-RT RIC) supports network-wide policy generation, enrichment, and orchestration at a slower timescale, whereas the Near-Real-Time RIC (Near-RT RIC) performs latency-sensitive radio control using live network measurements \cite{Bao26_CM}. This timescale separation naturally leads to distinct roles for AI across the RIC hierarchy. At the Non-RT RIC, AI can exploit historical and network-wide observations to predict network evolution and configure high-level control policies or criteria accordingly \cite{Bao26_CM}. At the Near-RT RIC, AI can use instantaneous network measurements to solve fast radio-control problems, such as resource allocation and power control. These complementary roles make the O-RAN control hierarchy a natural platform for adaptive AI-driven RAN operation.

Despite the increasing use of AI in RAN control, most existing solutions remain specialized to a predefined operating scenario and optimization task \cite{Dai25_TMC,Wu21_JSAC,Abedin22_TVT,Ju22_TWC,Tsampazi25_TMC,Noh24_CL}. Learning-based controllers are commonly trained and evaluated under a prescribed range of user densities, traffic patterns, mobility conditions, and radio environments. Therefore, their performance can deteriorate when the deployed network departs from these conditions \cite{Shen21_JSAC,Noh25_TVT}. In addition, as RANs support increasingly heterogeneous services, a single fixed optimization objective is no longer sufficient, since the objective and service constraints evolve with changing operational requirements \cite{Zappone19_TCOM,Wu22_COMST,Tsampazi25_TMC}. Existing approaches, however, typically embed a particular objective and constraint configuration into the optimization formulation or training loss, making each controller closely tied to the problem for which it was designed. An AI-native RAN therefore requires a control framework that can adapt not only to changing network environments, but also to different optimization problems presented after deployment.

\subsection{Related Works}

\paragraph{O-RAN control} The O-RAN RIC serves as a programmable control layer, enabling closed-loop radio control across multiple timescales. Its practical feasibility has been demonstrated through operational O-RAN platforms for training, deploying, and evaluating learning-based xApps~\cite{Polese23_TMC}, together with orchestration frameworks that coordinate control functions across RIC timescales~\cite{doro24_TMC}. Building on this programmability, RIC-based solutions have been developed for a wide range of radio-control tasks~\cite{Polese24_JSAC}, including traffic steering and handover management~\cite{Lacava24_TMC}, resource scheduling in virtualized RANs~\cite{Apostolakis24_JSAC}, and network slicing under service-level agreements~\cite{Dai25_TMC, Abedin22_TVT}. Despite these advances, most existing studies develop and evaluate their controllers under a prescribed operating environment characterized by a limited range of user densities, traffic patterns, mobility conditions, and radio configurations. As a result, the control logic is often closely coupled to the environment considered during design or training, and its performance can degrade when the deployed network deviates from these conditions~\cite{Tsampazi25_TMC,Noh25_TVT}. Thus, although O-RAN provides the programmability required to adapt network control after deployment, achieving robust operation across heterogeneous and evolving network environments remains an open challenge.

\paragraph{Environment-adaptive optimization} Learning-based optimization replaces repeated numerical optimization with mappings from network observations to control actions, enabling adaptation to variations in channel, traffic, and topology~\cite{Zappone19_TCOM}. Recent approaches improve generalization across operating conditions through graph-based parameterizations~\cite{Shen21_JSAC, Eisen20_TSP}, model-based deep unfolding~\cite{Chowdhury21_TWC}, and reinforcement learning for temporally coupled control~\cite{Naderializadeh21_TWC, Mei21_TCOM, AyalaRomero22_TMC}. Constrained learning further incorporates service requirements and can adapt constraint margins as the network conditions change~\cite{Wu21_JSAC, Naderializadeh23_TSP}. However, these methods generally keep the optimization criterion fixed during training: the objective, constrained metrics, and optimization directions remain predefined. Thus, a policy may generalize across network environments while still being specialized to a single optimization problem, and changes in the requested problem typically require transfer learning or retraining~\cite{Nguyen22_PROC}.

\paragraph{LLM-based control} Large language models (LLMs) have recently been applied to telecom standards, configuration, and network management~\cite{Zhou25_COMST, Maatouk24_CM, Chen24_WC,Park26_TVT}. Intent-based approaches use natural language to express desired network behavior and translate it into executable configurations~\cite{Zhou25_WC, Leivadeas23_COMST}, while agentic frameworks employ LLMs for planning and coordination of existing network functions~\cite{Xu24_WC, Jiang24_WC}. At the RAN level, LLM-hRIC provides high-level guidance to a reinforcement-learning controller across RIC timescales~\cite{Bao26_CM}, LLM-based resource allocation adapts control decisions to changing wireless environments~\cite{Noh25_TVT}, and multimodal models incorporate non-textual network observations together with language instructions~\cite{Yang25_VTM}. These studies demonstrate that language models can influence network control, but how to dynamically construct and adapt the numerical loss function according to changing operator requirements and network conditions remains unexplored.

\subsection{Challenges}
The existing studies reveal a fundamental gap: a unified O-RAN RIC architecture capable of adapting to both dynamic network conditions and changing operational requirements after deployment has yet to be established. Existing controllers are typically designed for either a prescribed operating environment or a fixed optimization criterion, limiting their ability to accommodate both types of change within a single deployed framework. This calls for a common RIC architecture that can adapt its control criterion and radio decisions to evolving network context and requirements without retraining a separate policy for each case. These observations motivate the following research question:
\begin{tcolorbox}[colframe=black, colback=white, height=1.5cm, boxrule=0.4mm]
\begin{center}
\vspace{-0.137cm}
\textit{\textbf{How can an O-RAN controller adapt to evolving network conditions and changing operational goals without retraining?}}
\end{center}
\end{tcolorbox}

\subsection{Contributions}

This paper proposes multimodal large language model (MLLM)-guided constrained optimization for RAN intelligent control (MLLM-coRIC), a requirement-adaptive O-RAN control framework that treats the optimization problem itself as a runtime input rather than a fixed design-time specification. At the Non-RT RIC, an operator request is compiled into a numerical specification of the objective and constraints, and the corresponding control loss is constructed using the predicted network context. The MLLM then refines the loss coefficients from measured objective and constraint outcomes instead of directly generating radio actions. At the Near-RT RIC, the resulting loss conditions a common numerical executor for joint resource and power allocation. This separation enables the same deployed controller to accommodate changes in both network conditions and operator-specified optimization problems without per-request retraining.

\begin{figure*}[t]
\centering
\includegraphics[width=\textwidth]{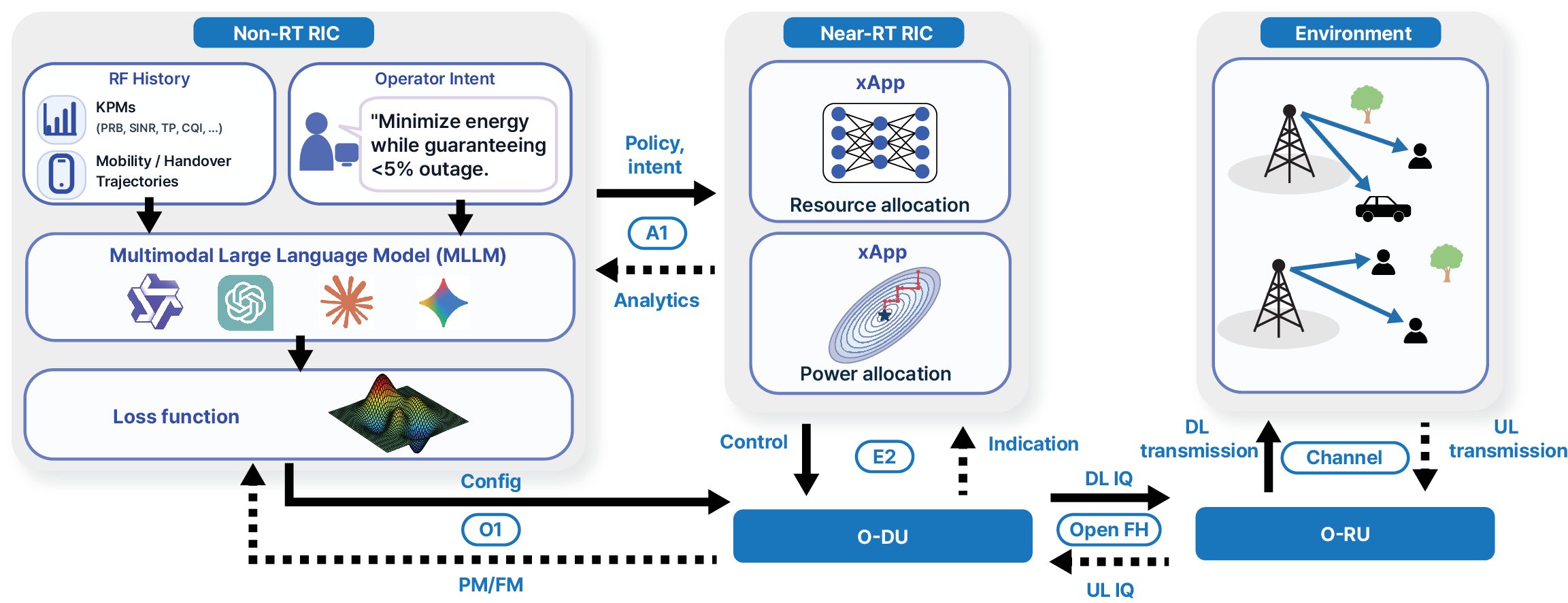}
\caption{Language-guided two-timescale O-RAN control. The Non-RT rApp compiles the operator's objective into a verified loss and forecasts per-cell demand; the numeric Near-RT xApp sets per-cell resource and power targets.}
\label{fig:arch}
\end{figure*}

The main contributions are summarized as follows:
\begin{itemize}[leftmargin=*]
    \item MLLM-coRIC is proposed as a requirement-adaptive hierarchical O-RAN control framework for joint resource and power allocation under changing network conditions and operator requirements. Unlike conventional task-specific controllers designed for a predefined objective and operating scenario, MLLM-coRIC jointly exploits the operator's natural-language specification and radio-frequency (RF)-derived network context as multimodal inputs. The hierarchical controller then solves the requested resource and power allocation problem under the observed network condition without requiring a dedicated optimization algorithm or task-specific policy to be redesigned or retrained for each request. Consequently, MLLM-coRIC extends RAN intelligence beyond solving a predefined task by enabling the optimization problem itself to be configured at runtime, allowing a single deployed framework to support heterogeneous operator-defined requirements across changing network conditions.

    \item At the Non-RT RIC, MLLM-coRIC exploits the reasoning capability of the MLLM to design the control loss from a holistic network perspective. By jointly interpreting the operator requirement, predicted network evolution, and measured optimization outcomes, the MLLM evaluates the relative importance of the objectives and constraints. Their penalty weights are then iteratively refined through a closed-loop process. The MLLM therefore operates at the slow timescale as a loss designer, embedding both the predicted network evolution and operator requirements into the optimization criterion executed by the fast controller.
    
    \item A loss-conditioned hybrid executor is developed for the Near-RT RIC to translate the synthesized optimization criterion into fast radio actions. The executor combines a learned resource allocation policy for ramp-constrained provisioning with a model-based interference-aware power controller, exploiting learning for sequential adaptation and analytical optimization for instantaneous power coordination. This hybrid design preserves low-latency execution while maintaining consistent optimization behavior under the requirement-dependent loss.

    \item A multi-cell O-RAN evaluation environment is implemented by integrating the CARLA urban mobility simulator with the Sionna RT ray-tracing-based wireless propagation simulator. Extensive evaluations under varying user loads and operator-specified objectives, constraints, and thresholds demonstrate that MLLM-coRIC can adapt a single deployed controller to different network conditions and optimization problems while satisfying the corresponding service requirements without retraining.
\end{itemize}

\section{System Model and Problem Formulation}
\label{sec:sys}

\subsection{Preliminary: O-RAN Control Hierarchy}
The O-RAN control hierarchy exposes different information granularities and control timescales across the Non-RT RIC and Near-RT RIC, as shown in Fig.~\ref{fig:arch}. This distinction determines which part of the end-to-end RAN control problem can be handled at each layer.

\paragraph{Non-RT RIC} The Non-RT RIC operates on a timescale of seconds or longer and has access to network-wide information aggregated over time rather than instantaneous per-physical resource block (PRB) scheduling states. It therefore provides slow-timescale orchestration by processing historical RF measurements and operator objectives and delivering policies or enrichment information to the Near-RT RIC through the A1 interface.

\paragraph{Near-RT RIC} The Near-RT RIC operates at a faster timescale and receives live cell- and user equipment (UE)-level measurements through the E2 interface. In this work, it performs cell-level resource and power control by determining the provisioned PRB budget $B_c(t)$ and the common per-PRB transmit power $P_c(t)$ for each cell, while leaving instantaneous UE-level PRB scheduling to the O-RAN distributed unit (O-DU).

\subsection{Communication Model}
Consider a network of $C$ co-channel cells indexed by $\mathcal{C}=\{1,\ldots,C\}$, where each cell provides $N_{\rm PRB}$ PRBs of bandwidth $W_{\rm PRB}$. At control epoch $t$, the network serves a time-varying set of UEs $\mathcal{U}_t$.

\paragraph{Channel, traffic, and load} Let $h_{u,c}(t)$ denote the channel from cell $c$ to UE $u$, whose power gain $|h_{u,c}(t)|^2$ comprises distance-dependent path loss and log-normal shadowing and is observed through the UE's reference-signal received power/quality (RSRP/RSRQ) measurements. UE $u\in\mathcal{U}_t$ has demand $d_u(t)$, fixed at the nominal rate of the service it draws upon attachment. Let $a_{u,c}(t)\in\{0,1\}$ denote the association indicator, where $a_{u,c}(t)=1$ if UE $u$ is associated with cell $c$ at epoch $t$, and $\sum_{c\in\mathcal{C}}a_{u,c}(t)=1$ for all $u$ and $t$. Strongest-channel association is assumed, i.e., $a_{u,c}(t)=\mathbf{1}\!\left\{c=\operatorname*{arg\,max}_{j\in\mathcal{C}}|h_{u,j}(t)|^2\right\}$. The offered load at cell $c$ is represented by
\begin{align}
\label{eq:load}
\lambda_c(t)=\sum_{u\in\mathcal{U}_t}a_{u,c}(t)d_u(t).
\end{align}

\paragraph{Resource and power allocation variables} At each control epoch $t$, the Near-RT RIC determines the provisioned PRB budget $B_c(t)\in\{B_{\min},\ldots,N_{\rm PRB}\}$ and the per-PRB transmit power $P_c(t)\in[P_{\min},P_{\max}]$ for each cell $c\in\mathcal{C}$. The variable $B_c(t)$ specifies the number of PRBs provisioned at cell $c$, whereas $P_c(t)$ specifies the transmit power applied to each utilized PRB. Increasing $B_c(t)$ increases the available radio capacity and the associated provisioning power consumption, while changing $P_c(t)$ affects both the desired received signal power and the inter-cell interference. The joint selection of $(B_c(t),P_c(t))$ is referred to as resource and power allocation.

\paragraph{Communication metric} Let $q_{u,c}(t)\in\mathbb{Z}_{\geq0}$ denote the number of PRBs allocated by cell $c$ to UE $u$ at epoch $t$. A UE can receive resources only from its associated cell, i.e., $q_{u,c}(t)\leq a_{u,c}(t)B_c(t)$, and the aggregate allocation at each cell satisfies $\sum_{u\in\mathcal{U}_t}q_{u,c}(t)\leq B_c(t)$. The PRB utilization ratio of cell $c$ is defined as
\begin{align}
\label{eq:prb_utilization}
\omega_c(t)=\frac{1}{N_{\rm PRB}}\sum_{u\in\mathcal{U}_t}q_{u,c}(t).
\end{align}
Under universal frequency reuse, the effective signal-to-interference-plus-noise ratio (SINR) of UE $u$ when served by cell $c$ is
\begin{align}
\label{eq:sinr}
\gamma_{u,c}(t)=\frac{P_c(t)|h_{u,c}(t)|^2}{N_0W_{\rm PRB}+\sum\limits_{j\in\mathcal{C}\setminus\{c\}}\omega_j(t)P_j(t)|h_{u,j}(t)|^2},
\end{align}
where $N_0$ is the receiver noise power spectral density. The utilization factor $\omega_j(t)$ captures the fraction of the PRB pool over which cell $j$ contributes inter-cell interference.

The spectral efficiency of UE $u$ when served by cell $c$ is $\mathrm{SE}_{u,c}(t)=\min\{\log_2(1+\gamma_{u,c}(t)),\mathrm{SE}_{\max}\}$. Accordingly, the delivered rate of UE $u$ is
\begin{align}
\label{eq:user_rate}
r_u(t)=W_{\rm PRB}\sum_{c\in\mathcal{C}}q_{u,c}(t)\mathrm{SE}_{u,c}(t).
\end{align}
The O-DU determines $\{q_{u,c}(t)\}$ for its associated UEs subject to the provisioned PRB budget $B_c(t)$. We assume a fixed demand-aware round-robin scheduler at the O-DU, such that UE-level scheduling is treated as an underlying network operation rather than a control variable of the RIC.

\paragraph{Power model} Each cell is modeled as operating in one of three states: DEEP, LIGHT, or ACT.
Sleeping cells consume only state-dependent idle power, whereas an active
cell incurs a fixed baseline cost, a bandwidth-provisioning cost proportional
to $B_c(t)$, and an RF cost determined by the actually utilized PRBs $L_c(t)$
and transmit power $P_c(t)$. Following the EARTH short-sleep model and
3GPP cell discontinuous transmission (DTX)~\cite{earth2011}, the power consumption can be represented by
\begin{align}
P_{\rm net}(t) &= \sum_{c\in\mathcal C} P_c(t), \\
P_c(t)
&=
\frac{E_{\rm tr}}{\Delta t}
\mathbf{1}\!\left[s_c(t)\neq s_c(t-1)\right]
\nonumber\\
&\hspace{-20pt}+
\begin{cases}
P_{\rm DEEP}, & s_c(t)=\mathrm{DEEP},\\
P_{\rm LIGHT}, & s_c(t)=\mathrm{LIGHT},\\
P_{\rm base}
+\alpha_b\frac{B_c(t)}{N_{\rm PRB}}
+P_{\rm rf}\frac{L_c(t)}{N_{\rm PRB}}
\frac{p_c(t)}{P_{\max}},
& s_c(t)=\mathrm{ACT}.
\end{cases}
\label{eq:power}
\end{align}
Thus, increasing $B_c(t)$ can keep a cell prewarmed at the cost of additional provisioning power, whereas the RF cost is incurred only by PRBs that actually carry traffic. The time-averaged network power is defined as
$P_{\rm tot}=\mathbb{E}_t[P_{\rm net}(t)]$.
The cell state $s_c(t)$ is decided by its traffic and provisioning level:
\begin{itemize}[leftmargin=*,itemsep=0pt,topsep=2pt]
\item $\text{ACT}$ -- the cell serves traffic or remains sufficiently provisioned as an active or prewarmed cell;
\item $\text{LIGHT}$ -- the cell serves no traffic but remains partially provisioned;
\item $\text{DEEP}$ -- the cell serves no traffic and is not maintained in a provisioned state.
\end{itemize}

\paragraph{Ramp constraint}
The provisioned PRB budget cannot change arbitrarily between consecutive control epochs because increasing the available bandwidth from a low-power state incurs a finite wake-up delay. Accordingly, the realized PRB budget is subject to
\begin{align}
-\delta_{\rm dn}^B
\le B_c(t)-B_c(t-1)
\le \delta_{\rm up}^B,
\label{eq:ramp}
\end{align}
where $\delta_{\rm up}^B$ and $\delta_{\rm dn}^B$ denote the maximum per-epoch increase and decrease in the PRB budget, respectively. In contrast, transmit power can be adjusted within each control epoch. This difference in actuation timescale motivates forecast-aided resource allocation (RA), whereas power allocation (PA) responds to the current channel and interference conditions.

\subsection{Problem Formulation}
This paper investigates joint resource and power allocation under operator-defined requirements. Specifically, the operator specifies the desired network objective and service constraints, which jointly define the optimization problem to be solved. Accordingly, the objective $J(t)$ and the constraints $\{g_j(t) \bowtie_j b_j\}$ are determined by the operator requirement and may vary across control episodes. Given these requirements, the RIC determines the per-cell provisioned PRB budgets $\mathbf B(t)=\{B_c(t)\}_{c\in\mathcal C}$ and per-PRB transmit powers $\mathbf P(t)=\{P_c(t)\}_{c\in\mathcal C}$. The resulting operator-defined resource and power allocation problem is formulated as
\begin{problem}{1}
    \begin{alignat}{3}
        & \bf{\Problem{1}}:
        && \min_{\{\mathbf B(t),\,\mathbf P(t)\}_{t=1}^{T_{\rm ep}}}
        && \mathbb{E}\!\left[\sum_{t=1}^{T_{\rm ep}}J(t)\right]
        \label{P1:objective}
        \\
        & && \mathrm{~~~~s.t.~}
        && \frac{\sum_{t=1}^{T_{\rm ep}}w_j(t)\,\mathbb{E}[g_j(t)]}
                {\sum_{t=1}^{T_{\rm ep}}w_j(t)}
        \ \bowtie_j\ b_j,\quad \forall j,
        \label{P1:service_constraint}
        \\
        & && &&
        B_c(t)\in\{B_{\min},\ldots,N_{\rm PRB}\},\quad \forall c,t,
        \label{P1:resource_allocation}
        \\
        & && &&
        P_c(t)\in[P_{\min},P_{\max}],\quad \forall c,t,
        \label{P1:power_allocation}
        \\
        & && &&
        -\delta_{\rm dn}^{B}\leq B_c(t)-B_c(t-1)\leq\delta_{\rm up}^{B},
        \quad \forall c,t.
        \label{P1:ramp_constraint}
    \end{alignat}
\end{problem}
Constraint \eqref{P1:service_constraint} specifies the operator-defined time-averaged service requirements, where $w_j(t)$ denotes the population over which metric $g_j(t)$ is averaged: $w_j(t)=1$ for a per-cell metric such as network power, and $w_j(t)=|\mathcal U_t|$ for a per-UE metric such as the quality-of-service (QoS) violation rate, whose averaging population varies with the number of active UEs. Constraints \eqref{P1:resource_allocation} and \eqref{P1:power_allocation} define the feasible ranges of the provisioned PRB budget and transmit power, respectively, while \eqref{P1:ramp_constraint} limits the per-epoch variation of the provisioned PRB budget.

Problem $\Problem{1}$ is, in general, a mixed-integer, non-convex, and temporally coupled stochastic control problem. The discrete PRB budgets introduce integer-valued decisions, while the transmit powers across neighboring cells are coupled through inter-cell interference in the effective SINRs. Moreover, the provisioned PRB budgets constrain the O-DU scheduling results, which in turn affect the achievable rates and PRB utilization, leading to a non-smooth dependence of the network performance on the control variables. The ramp constraint further couples the resource allocation decisions across consecutive epochs, while the traffic, mobility, and channel conditions evolve stochastically over time. Sec.~\ref{sec:method} develops a hierarchical O-RAN solution to address these coupled decision processes across the Non-RT and Near-RT control timescales.

\section{Proposed Language-Guided Hierarchical Resource and Power Allocation}
\label{sec:method}
\subsection{Overview: Hierarchical Decomposition of \texorpdfstring{$\Problem{1}$}{P1}}
\begin{figure*}[t]
\centering
\includegraphics[width=\textwidth]{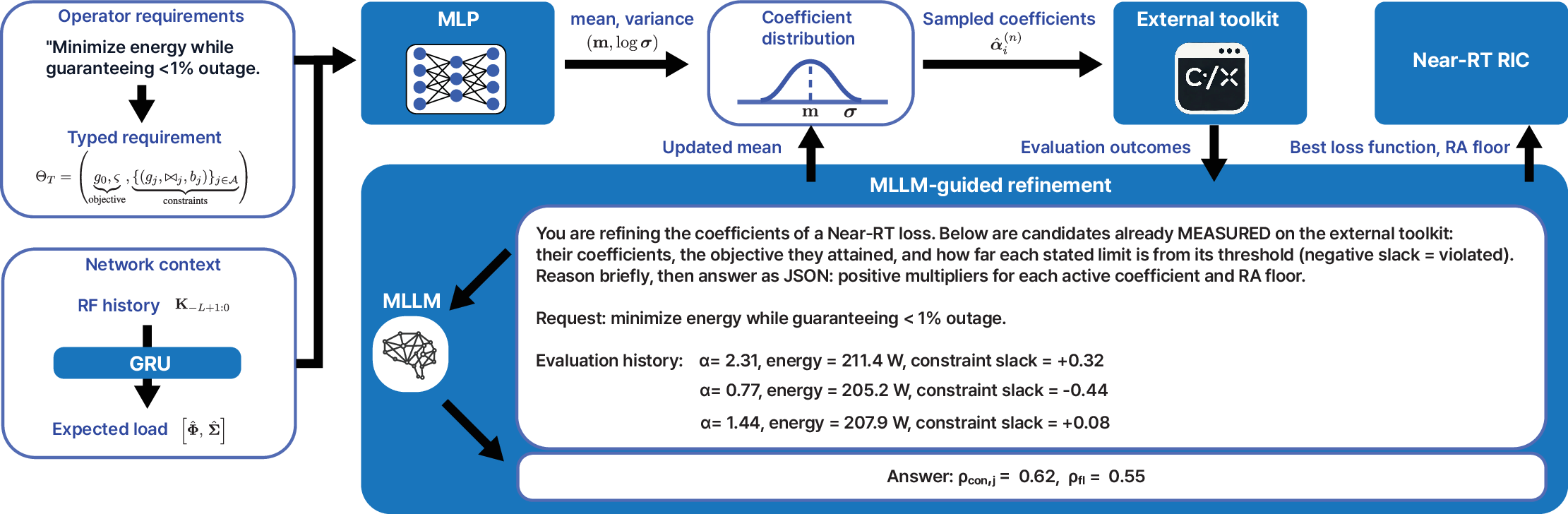}
\caption{The Non-RT rApp's refinement loop, with the prompt the MLLM receives and the answer it returns. A trained head turns the requirement and the radio state into a distribution over the loss coefficients; candidates drawn from it are scored by the network-evaluation toolkit; the MLLM reads the measured table and answers one relative multiplier per term, which recenters the distribution for the next round. Only the finalized $\boldsymbol{\alpha}^{(N_{\rm ref})}$ and $\underline B_{c,t}^{(N_{\rm ref})}$ leave the loop.}
\label{fig:nonrtloop}
\end{figure*}

The objective $J(t)$ and service constraints $\{g_j(t)\bowtie_j b_j\}$ in $\Problem{1}$ are not fixed a priori, but are specified by the operator through a natural-language request. At each Non-RT interval, the Non-RT RIC compiles the request into a typed numerical problem specification, forecasts the future cell loads, and constructs a requirement-dependent control loss by adapting the relative penalties of the objective and constraints to the predicted network context and measured optimization outcomes. The detailed flowchart of the Non-RT RIC is illustrated in Fig.~\ref{fig:nonrtloop}. The resulting loss specification and load forecast are delivered to the Near-RT RIC through the A1 interface. At each Near-RT epoch, the Near-RT RIC combines this A1 enrichment with live E2 measurements to determine the per-cell PRB budgets $\mathbf B(t)$ and transmit powers $\mathbf P(t)$.

\subsection{Non-RT rApp: Requirement Compilation, Forecasting, and Measured Loss Synthesis}
\label{sec:losssynth}

\paragraph{Traffic-load forecasting} Before each control episode, the Non-RT RIC predicts the future offered load of each cell from a recent history of network-wide RF measurements. Let $t=0$ denote the beginning of the control episode, with $t=1,\ldots,T_{\rm ep}$ denoting the subsequent control epochs, and let $\tau\in\{-L+1,\ldots,0\}$ index the preceding $L$ RF observations. At each observation $\tau$, the RF measurements associated with cell $c$ are summarized as
\begin{align}
\label{eq:forecast_feat}
\mathbf k_{c,\tau}=\left[\lambda_c(\tau),\,n_c(\tau),\,\bar{\gamma}_c(\tau),\,\gamma_c^{5\%}(\tau),\,\bar{\mathbf p}_{c,\tau},\,\bar{\mathbf v}_{c,\tau},\,\bar{\mathbf g}_{c,\tau}\right],
\end{align}
where $\lambda_c(\tau)$ and $n_c(\tau)$ denote the offered load and the number of associated UEs, respectively, $\bar{\gamma}_c(\tau)$ and $\gamma_c^{5\%}(\tau)$ denote the average and $5$th-percentile serving-link SINRs of the associated UEs, $\bar{\mathbf p}_{c,\tau}$ and $\bar{\mathbf v}_{c,\tau}$ denote their average position and velocity relative to cell $c$, and $\bar{\mathbf g}_{c,\tau}\in\mathbb{R}^{C}$ contains their average channel gains toward all cells in $\mathcal C$. The SINR statistics capture the overall and cell-edge link quality, while the mobility and cross-cell channel measurements provide early indications of traffic migration before the associated traffic is transferred to a neighboring cell. The per-cell features are concatenated as $\hat{\mathbf{k}}_\tau=[\mathbf k_{1,\tau},\ldots,\mathbf k_{C,\tau}]$, and the RF history preceding the control episode is denoted by $\mathbf K_{-L+1:0}=\{\hat{\mathbf{k}}_\tau\}_{\tau=-L+1}^{0}$.

A standard gated recurrent unit (GRU)-based forecaster~\cite{cho2014learning} maps the recent RF history to the mean and uncertainty of the expected loads:
\begin{align}
\label{eq:forecast}
\left[\hat{\bm\Phi},\,\hat{\bm\Sigma}\right]
=g_{\theta_{\rm f}}\!\left(\mathbf K_{-L+1:0}\right),
\end{align}
where $\mathbf K_{-L+1:0}$ denotes the RF history observed before the control episode, $\hat{\bm\Phi}=[\hat{\phi}_{c,t}]_{c\in\mathcal C,\,t=1:T_{\rm ep}}$ contains the predicted offered loads, and $\hat{\bm\Sigma}=[\hat{\sigma}_{c,t}]_{c\in\mathcal C,\,t=1:T_{\rm ep}}$ contains the corresponding predictive uncertainties. Specifically, $\hat{\phi}_{c,t}$ predicts the offered load $\lambda_c(t)$ of cell $c$ at future control epoch $t$. The model parameters $\theta_{\rm f}$ are trained using a heteroscedastic Gaussian negative log-likelihood.

\paragraph{Requirement} The natural-language request, including the objective and constraints, is denoted by $o_T$. The MLLM interprets $o_T$ and compiles the explicitly stated optimization objective and service constraints into the typed requirement
\begin{align}
\label{eq:compile}
\Theta_T
=\left(
\underbrace{g_0,\varsigma}_{\text{objective}},
\underbrace{\{(g_j,\bowtie_j,b_j)\}_{j\in\mathcal A}}_{\text{constraints}}
\right),
\end{align}
where $g_0$ denotes the requested objective metric, $\varsigma\in\{-1,+1\}$ its optimization direction, and each $(g_j,\bowtie_j,b_j)$ specifies the metric, inequality direction, and threshold of an operator-stated constraint. This MLLM-based compilation determines the structure of the requested optimization problem, while the relative weighting between the objective and constraints is adapted to the network condition through the loss-synthesis process described next.

\paragraph{Loss construction}
To translate the operator requirement $\Theta_T$ into a form executable by the Near-RT RIC, the requested objective and active constraints are expressed through the loss function
\begin{align}
\label{eq:loss_form}
\mathcal L_T
=
\underbrace{\varsigma\,g^{\star}}_{\text{objective}}
+
\sum_{j\in\mathcal A}
\underbrace{
\alpha_j
\big[\varsigma_j(g_j-b_j)\big]_{+}
}_{\text{constraint }j},
\end{align}
where $\varsigma_j\in\{-1,+1\}$ sets the direction of constraint $j$ such that a positive value of $\varsigma_j(g_j-b_j)$ indicates a constraint violation. The requirement $\Theta_T$ determines the objective, active constraint set $\mathcal A$, and corresponding thresholds, while
\begin{align}
\label{eq:alpha_target}
\boldsymbol{\alpha}
=
\{\alpha_j\}_{j\in\mathcal A}
\end{align}
determines the relative penalty assigned to each active constraint.

A fixed $\boldsymbol{\alpha}$ may not provide consistent constraint satisfaction across different network conditions, since the required penalty strengths vary with the expected traffic and radio states. The Non-RT RIC therefore adapts $\boldsymbol{\alpha}$ to the current network context. A multilayer perceptron (MLP) first initializes the coefficients from $\Theta_T$, the current RF state $\mathbf k_0$, and the forecast $(\hat{\bm\Phi},\hat{\bm\Sigma})$, after which the MLLM refines them using the measured objective and constraint outcomes from the external network-evaluation loop. The refined $\boldsymbol{\alpha}$, together with the floor $\underline B_{c,t}$, is delivered to the Near-RT RIC to condition the resource allocation policy and parameterize the power allocation objective.

\paragraph{Initial coefficients from an MLP}
The Non-RT RIC first constructs a context-dependent distribution over the constraint coefficients using a trained MLP $h_{\bm\theta_{\rm init}}$:
\begin{align}
\label{eq:losshead}
(\mathbf m,\log\bm\sigma)
=
h_{\bm\theta_{\rm init}}\!\left(
\Theta_T,\mathbf k_0,\hat{\bm\Phi},\hat{\bm\Sigma}
\right),
\end{align}
where $\mathbf m,\bm\sigma\in\mathbb R^{|\mathcal A|}$ parameterize the mean and standard deviation in the log-coefficient domain. Candidate coefficient vectors are then sampled as
\begin{align}
\label{eq:head_dist}
\log \hat{\boldsymbol{\alpha}}_i^{(0)}
=
\mathbf m^{(0)}
+
\bm\sigma\odot\bm\epsilon_i,
\qquad
\bm\epsilon_i
\sim
\mathcal N(\mathbf 0,\mathbf I),
\end{align}
where $i$ indexes the sampled candidates. From \eqref{eq:head_dist}, $N_\text{s}$ samples are generated.

\paragraph{MLLM-guided refinement} 
The MLP provides a context-dependent initialization of the loss-coefficient distribution, from which candidate coefficient vectors are sampled and evaluated using the external network-evaluation toolkit. The toolkit returns the achieved objective and constraint outcomes for each candidate, and the MLLM uses these measured results to determine relative coefficient updates for the next refinement round. The updated coefficients recenter the sampling distribution, and this evaluation--refinement process is repeated iteratively. In parallel, the MLLM adjusts the forecast-guided floor $\underline B_{c,t}$ to tighten or relax the minimum resource provisioning passed to the Near-RT RIC.

The refinement starts from the MLP's proposal $\hat{\boldsymbol{\alpha}}_i^{(0)}$ in \eqref{eq:head_dist}. At round $n$, the coefficients $\hat{\boldsymbol{\alpha}}_i^{(n)}$ are applied to \eqref{eq:loss_form} and the toolkit returns
\begin{align}
\label{eq:tool_feedback}
\mathbf y_i^{(n)}
=\left(g_{0,i}^{(n)},\ \left\{\varsigma_j\big(g_{j,i}^{(n)}-b_j\big)\right\}_{j\in\mathcal A}\right),
\end{align}
where $g_{0,i}^{(n)}$ and $g_{j,i}^{(n)}$ denote the measured objective value and the value of constrained metric $j$ returned by the toolkit, respectively. The coefficient--feedback pairs accumulate into the evaluation history $\mathcal H^{(n)}=\{\{(\hat{\bm\alpha}_i^{(m)},\mathbf y_i^{(m)})\}_{i=0}^{N_\text{s}-1}\}_{m=0}^{n}$. The MLLM reads this history and returns the next coefficients,
\begin{align}
\label{eq:alpha_refine}
\boldsymbol{\alpha}^{(n+1)}=\mathcal R_{\bm\psi}\big(\mathcal H^{(n)};\,\Theta_T,\mathbf k_0,\hat{\bm\Phi},\hat{\bm\Sigma}\big),
\end{align}
where $\mathcal R_{\bm\psi}$ denotes the MLLM-based refinement mapping, realized by the forward pass detailed below.

\textit{1) Tokenization and embedding:} The requirement, the network context, and the measured outcomes of the current round are serialized into a token sequence $\mathbf u=\mathcal T(\Theta_T,\mathbf k_0,\hat{\bm\Phi},\hat{\bm\Sigma},\mathcal H^{(n)})\in\mathcal V^{N_{\rm tok}}$ over the vocabulary $\mathcal V$, where $\mathcal T$ denotes the prompt template. Each token is embedded through $\mathbf E\in\mathbb R^{|\mathcal V|\times d}$ and stacked into the latent representation matrix
\begin{align}
\label{eq:mllm_embed}
\mathbf Y=\big[\mathbf E_{u_1}\ \mathbf E_{u_2}\ \cdots\ \mathbf E_{u_{N_{\rm tok}}}\big]^{\sf T}\in\mathbb R^{N_{\rm tok}\times d}.
\end{align}

\textit{2) Backbone:} $\mathbf Y$ is processed by $L$ sequential transformer blocks. In each block, $H$ attention heads operate in parallel. Let $f_{\rm ln}(\cdot)$ denote layer normalization. With $\bar{\mathbf Y}=f_{\rm ln}(\mathbf Y)$, the $h$th head forms the query, key, and value matrices
\begin{align}
\label{eq:mllm_qkv}
\mathbf Q_h&=\bar{\mathbf Y}\mathbf W_{q,h},\\
\mathbf K_h&=\bar{\mathbf Y}\mathbf W_{k,h},\\
\mathbf V_h&=\bar{\mathbf Y}\mathbf W_{v,h},
\end{align}
where $\mathbf W_{q,h},\mathbf W_{k,h},\mathbf W_{v,h}\in\mathbb R^{d\times d_h}$ are trainable projection matrices and $d_h=d/H$. The output of the $h$th attention head is represented by
\begin{align}
\label{eq:mllm_attn}
\mathbf Y_{{\rm att},h}=f_{\rm smax}\!\left(\frac{\mathbf Q_h\mathbf K_h^{\sf T}}{\sqrt{d_h}}+\mathbf M\right)\mathbf V_h,
\end{align}
where $f_{\rm smax}(\cdot)$ denotes the row-wise softmax and $\mathbf M\in\mathbb R^{N_{\rm tok}\times N_{\rm tok}}$ is the causal mask, with $[\mathbf M]_{i,j}=0$ for $j\leq i$ and $-\infty$ otherwise. The head outputs are concatenated and projected as
\begin{align}
\label{eq:mllm_concat}
\mathbf Y_{\rm att}=\big[\mathbf Y_{{\rm att},1}\ \mathbf Y_{{\rm att},2}\ \cdots\ \mathbf Y_{{\rm att},H}\big]\mathbf W_o,
\end{align}
where $\mathbf W_o\in\mathbb R^{d\times d}$ is the output projection matrix. The output of the first transformer block is given by
\begin{align}
\label{eq:mllm_block}
\mathbf Z_1=\mathbf Y+\mathbf Y_{\rm att}+f_{\rm ffn}\!\left(f_{\rm ln}\!\left(\mathbf Y+\mathbf Y_{\rm att}\right)\right),
\end{align}
where $f_{\rm ffn}(\cdot)$ denotes the position-wise feed-forward network. Repeating the same block operation yields the final hidden representation $\mathbf Z=\mathbf Z_L\in\mathbb R^{N_{\rm tok}\times d}$.

\textit{3) Autoregressive decoding:} The final transformer representation is decoded autoregressively through the pretrained language-modeling (LM) head. At decoding step $\ell$, let $\mathbf z_\ell\in\mathbb R^d$ denote the final-layer hidden representation at the current decoding position, conditioned on the input sequence and the previously generated tokens. Under greedy decoding, the next token is selected as
\begin{align}
\label{eq:mllm_head}
x_{\ell+1}
=\arg\max_{v\in\mathcal V}
\left[\mathbf W_{\rm lm}\mathbf z_\ell+\mathbf b_{\rm lm}\right]_v,
\end{align}
where $\mathbf W_{\rm lm}\in\mathbb R^{|\mathcal V|\times d}$ and $\mathbf b_{\rm lm}\in\mathbb R^{|\mathcal V|}$ are the pretrained LM-head parameters. Repeated application of \eqref{eq:mllm_head} produces the output token sequence $\mathbf x_T=(x_{N_{\rm tok}+1},\ldots,x_{N_{\rm tok}+N_{\rm gen}})$. Since the output follows a fixed format, the generated sequence specifies the multiplicative factors applied to the coefficients in \eqref{eq:loss_form}, which are defined as
\begin{align}
\label{eq:alpha_llm}
\bm\rho^{(n)}
=\{\rho_{{\rm con},j}^{(n)}\}_{j\in\mathcal A}\cup\{\rho_{\rm fl}^{(n)}\},
\quad \rho\in \mathbb{R}^{+}.
\end{align}
Each $\rho_{{\rm con},j}^{(n)}$ is a multiplicative factor applied to the corresponding constraint coefficient, while $\rho_{\rm fl}^{(n)}$ scales the forecast-guided floor $\underline B_{c,t}$. Thus, the MLLM does not directly generate loss coefficients or Near-RT control actions, but only determines how the current coefficients and provisioning floor should be adjusted based on the measured outcomes.

The MLLM output $\bm\rho^{(n)}$ is applied multiplicatively to the best-performing candidate $\hat{\boldsymbol{\alpha}}_{i^\star}^{(n)}$, where $i^\star$ denotes the candidate with the best score according to $\mathbf y_i^{(n)}$. For each constraint $j\in\mathcal A$, the coefficient is updated as
\begin{align}
\label{eq:alpha_update}
\alpha_j^{(n+1)}
=
\hat{\alpha}_{i^\star,j}^{(n)}
\rho_{{\rm con},j}^{(n)}.
\end{align}
In parallel, the forecast-guided floor is also updated as
\begin{align}
\label{eq:floor_update}
\underline B_{c,t}^{(n+1)}
=
\underline B_{c,t}^{(n)} \rho_{\rm fl}^{(n)}.
\end{align}
The updated coefficient vector $\boldsymbol{\alpha}^{(n+1)}$ then recenters the log-domain sampling distribution for the next refinement round according to
\begin{align}
\label{eq:mean_update}
\mathbf m^{(n+1)}
=
\log \boldsymbol{\alpha}^{(n+1)}.
\end{align}
The candidate sampling, MLLM-guided refinement, and mean update are repeated for $N_{\rm ref}$ rounds, after which $\boldsymbol{\alpha}^{(N_{\rm ref})}$ and $\underline B_{c,t}^{(N_{\rm ref})}$ are delivered to the Near-RT RIC.

\subsection{Near-RT xApp: Loss-Conditioned Hybrid Executor and Training}
\label{sec:nearrt}

At each control epoch $t$, the Near-RT RIC translates the Non-RT guidance into resource and power allocation decisions using the current E2 measurements. The PRB budget is first determined based on the load forecast and requirement-dependent loss subject to the ramp constraint. Given the selected PRB budget, the transmit power is then optimized using the current channel and interference conditions. For cell $c$, the local state is defined as
\begin{align}
\label{eq:cell_feature}
\mathbf{x}_{c,t}
=
\big[
\lambda_c(t),\,
n_c(t),\,
\bar{\gamma}_c(t),\,
B_c(t-1),\,
P_c(t-1)
\big].
\end{align}

\paragraph{Resource allocation}
A compact MLP $f_{\bm\theta_{\rm RA}}$ maps the current network state, load forecast, and requirement specification to cell-wise latent representations,
\begin{align}
[\mathbf h_{1,t},\ldots,\mathbf h_{C,t}]
=
f_{\bm\theta_{\rm RA}}\!\left(
\{\mathbf x_{c,t}\}_{c\in\mathcal C},
\hat{\bm\Phi},
\hat{\bm\Sigma},
\Theta_T,
\boldsymbol{\alpha}
\right).
\end{align}
The budget head first determines a cell-specific provisioning range by producing
\begin{align}
\label{eq:budget_head}
\bar B_{c,t}
=
B_{\min}
+
(N_{\rm PRB}-B_{\min})
f_{\rm sig}\!\left(
\mathbf w_B^{\sf T}\mathbf h_{c,t}+b_B
\right),
\end{align}
where $\bar B_{c,t}\in[B_{\min},N_{\rm PRB}]$ represents the upper provisioning level suggested by the RA network. The forecast-guided floor $\underline B_{c,t}$ provides the corresponding minimum provisioning level.
Within this range, a small set of candidate budgets is constructed from the PRB demand observed at the preceding epoch. Let $\tilde B_{c,t}$ denote the measured PRB demand and let $\mathcal K\subset\mathbb R_{+}$ be a finite set of scaling factors. For each $\kappa\in\mathcal K$, the candidate budget is given by
\begin{align}
\label{eq:budget_candidate}
B^{(\kappa)}_{c,t}
=
\max\!\left(
\underline B_{c,t},\,
\min\!\left(
\bar B_{c,t},
\kappa\,\tilde B_{c,t}
\right)
\right).
\end{align}
For each $\kappa\in\mathcal K$, the RA module constructs a candidate PRB budget $\mathbf B_t^{(\kappa)}$. Each candidate is passed to the PA module and evaluated under the requirement-dependent loss. The final resource and power allocation is determined jointly after evaluating all candidate budgets.

\paragraph{Power allocation}
For each candidate budget $\mathbf B_t^{(\kappa)}$, the corresponding transmit-power vector is obtained by solving
\begin{align}
\label{eq:pa_ctrl}
\mathbf P_t^{(\kappa)}
=
\operatorname*{arg\,min}_{\mathbf P\in[P_{\min},P_{\max}]^{C}}
\mathcal L^{\rm PA}_T
\!\left(
\mathbf B_t^{(\kappa)},\mathbf P
\right).
\end{align}
For each active constraint $j\in\mathcal A$, the normalized violation under candidate $\kappa$ is defined as
\begin{align}
\label{eq:violation}
v_j\!\left(\mathbf B_t^{(\kappa)},\mathbf P\right)
=
\Big[
\varsigma_j
\big(
g_j(\mathbf B_t^{(\kappa)},\mathbf P)-b_j
\big)
\Big]_+ .
\end{align}
The violation is evaluated through the Huber penalty
\begin{align}
\label{eq:huber}
\xi_\delta(v)
=
\begin{cases}
\tfrac{1}{2}v^2, & v\leq\delta,\\[2pt]
\delta\left(v-\tfrac{1}{2}\delta\right), & v>\delta,
\end{cases}
\end{align}
yielding the requirement-dependent PA loss
\begin{align}
\label{eq:nearrt_loss}
\mathcal L^{\rm PA}_T
\!\left(
\mathbf B_t^{(\kappa)},\mathbf P
\right)
=
\varsigma\,
g_0\!\left(
\mathbf B_t^{(\kappa)},\mathbf P
\right)
+
\sum_{j\in\mathcal A}
\alpha_j
\xi_\delta\!\left(
v_j\!\left(
\mathbf B_t^{(\kappa)},\mathbf P
\right)
\right).
\end{align}
For each $\kappa$, \eqref{eq:pa_ctrl} is solved to obtain $\mathbf P_t^{(\kappa)}$, and the resulting resource--power pair is evaluated using \eqref{eq:nearrt_loss}. The best candidate is then selected as
\begin{align}
\label{eq:budget_select}
\kappa_t^\star
=
\operatorname*{arg\,min}_{\kappa\in\mathcal K}
\mathcal L^{\rm PA}_T
\!\left(
\mathbf B_t^{(\kappa)},
\mathbf P_t^{(\kappa)}
\right),
\end{align}
and the final control decisions are given by
\begin{align}
\label{eq:joint_selection}
\mathbf B_t
=
\mathbf B_t^{(\kappa_t^\star)},
\qquad
\mathbf P_t
=
\mathbf P_t^{(\kappa_t^\star)}.
\end{align}

The PA problem in \eqref{eq:pa_ctrl} is solved approximately by block-coordinate descent over the cell transmit powers. Each block update minimizes \eqref{eq:nearrt_loss} with respect to one $P_c$ while keeping the remaining cell powers fixed, and the iterations terminate when the power allocation converges.

\paragraph{Training}
The RA network is trained offline using episodic rollouts over the admissible requirement space. At each epoch, the RA network determines the PRB budget $\mathbf B_t(\bm\theta_{\rm RA})$, while the corresponding transmit-power vector $\mathbf P_t$ is obtained from the model-based PA solver. The RA parameters are trained to minimize the requirement-dependent loss together with a regularization term on inter-epoch budget variations:
\begin{align}
\label{eq:ra_loss}
\min_{\bm\theta_{\rm RA}}\quad
\mathbb E \Bigg[
&\sum_{t=1}^{T_{\rm ep}}
\{
\widetilde{\mathcal L}_T
(
\mathbf B_t(\bm\theta_{\rm RA}),
\mathbf P_t;
\Theta_T,
\boldsymbol{\alpha}
)
\nonumber \\
&\hspace{30pt}+
\alpha_{\rm slew}
\|
\mathbf B_t(\bm\theta_{\rm RA})-\mathbf B_{t-1}
\|_2^2
\}\Bigg],
\end{align}
where $\widetilde{\mathcal L}_T$ is a differentiable surrogate of the loss in \eqref{eq:nearrt_loss}, and $\alpha_{\rm slew}$ controls the temporal variation of the PRB budget. During training, $(\Theta_T,\boldsymbol{\alpha})$ is sampled over its admissible range, and the rollout follows the ramp-constrained budget dynamics in \eqref{eq:ramp}. The PA solution $\mathbf P_t$ is treated as fixed when updating $\bm\theta_{\rm RA}$. This enables a single RA network to learn requirement-aware resource provisioning while accounting for future demand under the ramp constraint.

\section{Performance Evaluation}
\label{sec:sim}

\subsection{Evaluation Setup}
\label{sec:setup}

\begin{figure}[t]
\centering
\includegraphics[width=\columnwidth]{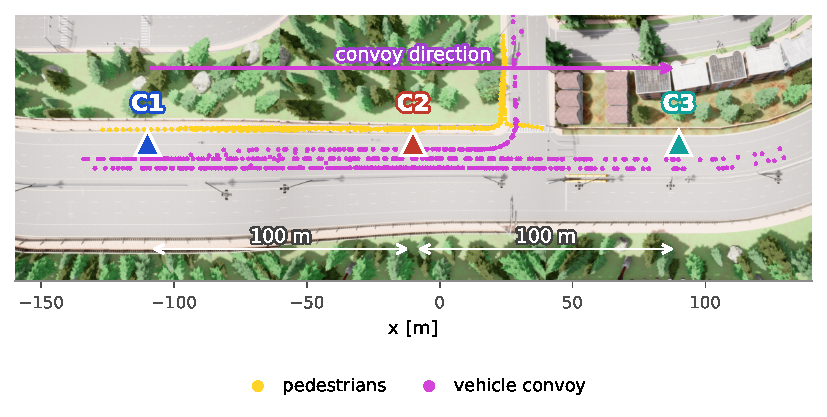}
\caption{The served corridor in CARLA Town05, seen from above. Triangles are the three cell sites and dots the recorded UE positions of one episode.}
\label{fig:town05map}
\end{figure}

\paragraph{Deployment and mobility}
The evaluation considers a three-cell deployment in CARLA Town05, where the cells are spaced
$100$\,m apart along a straight urban road, as shown in Fig.~\ref{fig:town05map}. In each
episode, a vehicle convoy moves
from C1 to C3 at approximately $30$\,km/h, creating a migrating traffic hotspot across
the cells. In addition, $10$ pedestrians remain locally distributed
around the cells and provide stationary background traffic. Each episode lasts $30$\,s. 

\paragraph{Traffic}
Each UE is assigned a service-dependent traffic demand upon attachment.
Pedestrians generate voice, web/social, or video traffic, whereas vehicles generate
vehicle-to-everything (V2X) safety, cooperative-driving, or sensor-sharing traffic following the 3GPP eV2X
requirements~\cite{ts22186}. The network load is varied by changing the convoy size
from $5$ to $25$ vehicles, resulting in heterogeneous and time-varying offered traffic
across the cells.

\paragraph{Radio and power}
The wireless channel follows the 3GPP urban micro (UMi) model at $3.5$\,GHz with log-normal
shadow fading. Each cell operates over $N_{\rm PRB}=133$ PRBs corresponding to an
approximately $50$\,MHz n78 channel, with a maximum transmit power of $40$\,dBm
per cell and a $256$-QAM spectral-efficiency cap. The cell power consumption follows
the EARTH micro base station (BS) model~\cite{earth2011} with the deep-, light-, and active-state
power model in \eqref{eq:power}. The Non-RT MLLM uses only the network-wide RF history as its radio-side input.

\paragraph{Baselines}
Three baseline methods are considered:
\begin{itemize}[leftmargin=*,itemsep=0pt,topsep=2pt]
\item Lyapunov drift-plus-penalty (DPP): PRB sizing via virtual-queue DPP control.
\item Greedy: PRB provisioning based on the current UE demands.
\item End-to-end deep reinforcement learning (DRL): model-free joint PRB and power control from the RF state.
\end{itemize}

\subsection{Problem-Solving Capability}
\label{sec:sweeps}

\begin{figure}[t]
\centering
\includegraphics[width=\columnwidth]{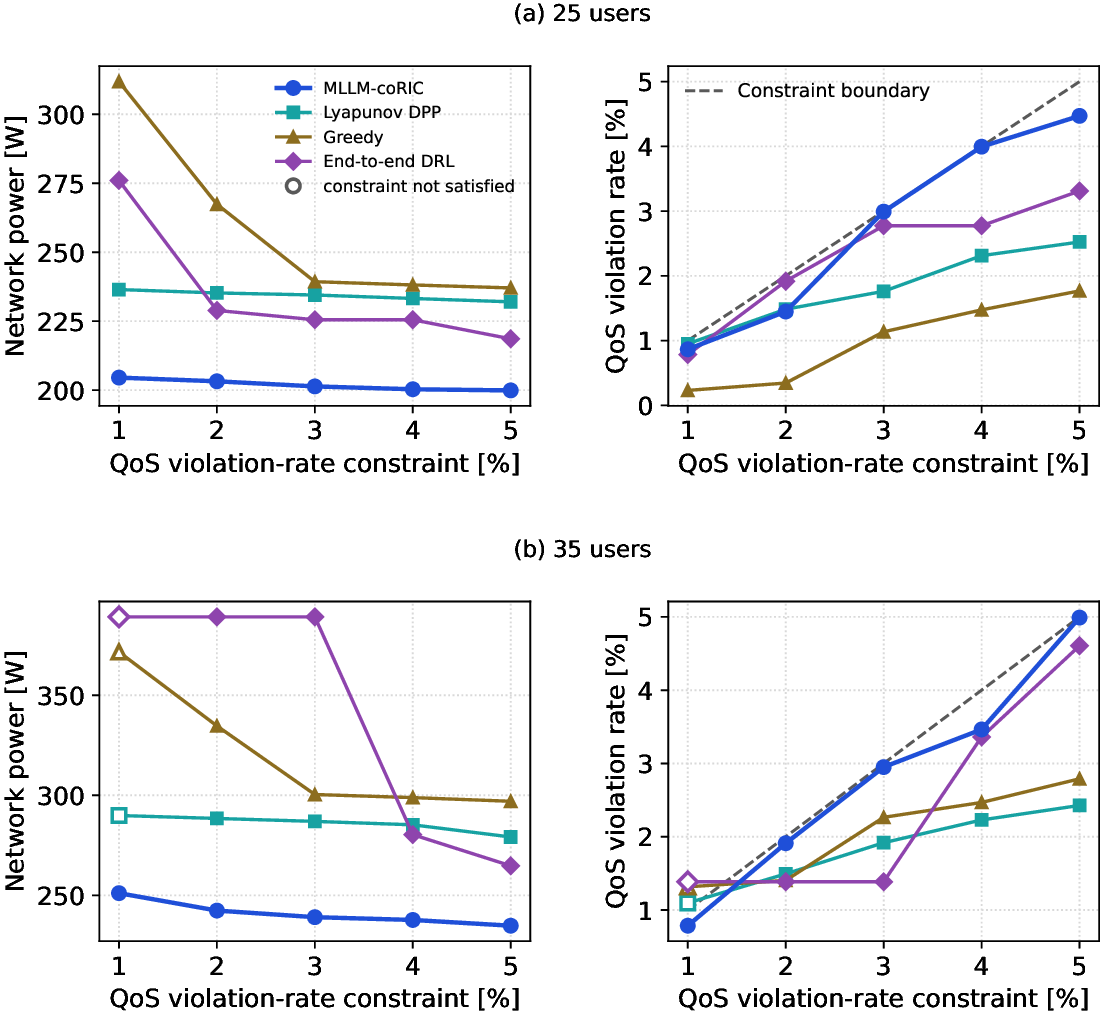}
\caption{Network power and QoS violation rate under the violation-rate constraint $\overline V$ ($\Problem{2}$): (a) $25$ users, (b) $35$ users. Dashed line: the constraint boundary $\mathrm{V}=\overline V$. Hollow markers mean that the method fails to satisfy the constraint.}
\label{fig:sw_out}
\end{figure}

This subsection first examines whether a single deployed MLLM-coRIC controller can accommodate different operator-defined constrained optimization problems without task-specific retraining. Let $\mathrm{V}$ denote the time-averaged QoS violation rate and $P_{\rm tot}$ the time-averaged network power, i.e.,
\begin{align}
\mathrm{V}
&=
\frac{\sum_{t=1}^{T_{\rm ep}}\sum_{u\in\mathcal U_t}
\mathbf{1}\left[r_u(t)<d_u(t)\right]}
{\sum_{t=1}^{T_{\rm ep}}|\mathcal U_t|},
\\
P_{\rm tot}&=
\frac{1}{T_{\rm ep}}
\sum_{t=1}^{T_{\rm ep}}
\sum_{c\in\mathcal C} P_c(t).
\end{align}
Then, two instances of $\Problem{1}$ are considered:
\begin{align} \Problem{2}:\ \min_{\{\mathbf{B}(t),\mathbf {P}(t)\}_{t=1}^{T_{\rm ep}}} P_{\rm tot} &\quad \text{s.t.}\quad \mathrm{V}\le\overline V, \label{eq:req_power}\\ 
\Problem{3}:\ \min_{\{\mathbf B(t),\mathbf P(t)\}_{t=1}^{T_{\rm ep}}} \mathrm{V} &\quad \text{s.t.}\quad P_{\rm tot}\le\overline P, \label{eq:req_outage} \end{align} 
subject to \eqref{P1:resource_allocation}--\eqref{P1:ramp_constraint}. The two requests exchange the objective and constraint, and therefore require different power--QoS operating points from the same radio-control framework.

Under $\Problem{2}$, MLLM-coRIC consistently operates near the stated QoS boundary while achieving the lowest network power, as shown in Fig.~\ref{fig:sw_out}. This behavior results from the requirement-conditioned loss synthesis: the Non-RT RIC adapts the coefficient associated with the QoS constraint based on the specified requirement, network context, and measured outcomes, preventing the controller from maintaining an unnecessarily conservative QoS margin when the constraint is relaxed. The resulting loss conditions the Near-RT executor, where forecast-guided resource allocation anticipates future demand and interference-aware power control coordinates transmit powers across cells. Together with the loss-aware budget search, this enables the controller to place radio resources where they are needed and approach the requested QoS boundary without unnecessary overprovisioning. Consequently, at $35$ users, the QoS violation rate of MLLM-coRIC increases from $0.79\%$ to $4.99\%$ as $\overline V$ is relaxed from $1\%$ to $5\%$, while MLLM-coRIC consistently achieves the lowest network power. It is also the only method that satisfies the stringent $1\%$ constraint, with the same trend observed at $25$ users.

\begin{figure}[t]
\centering
\includegraphics[width=\columnwidth]{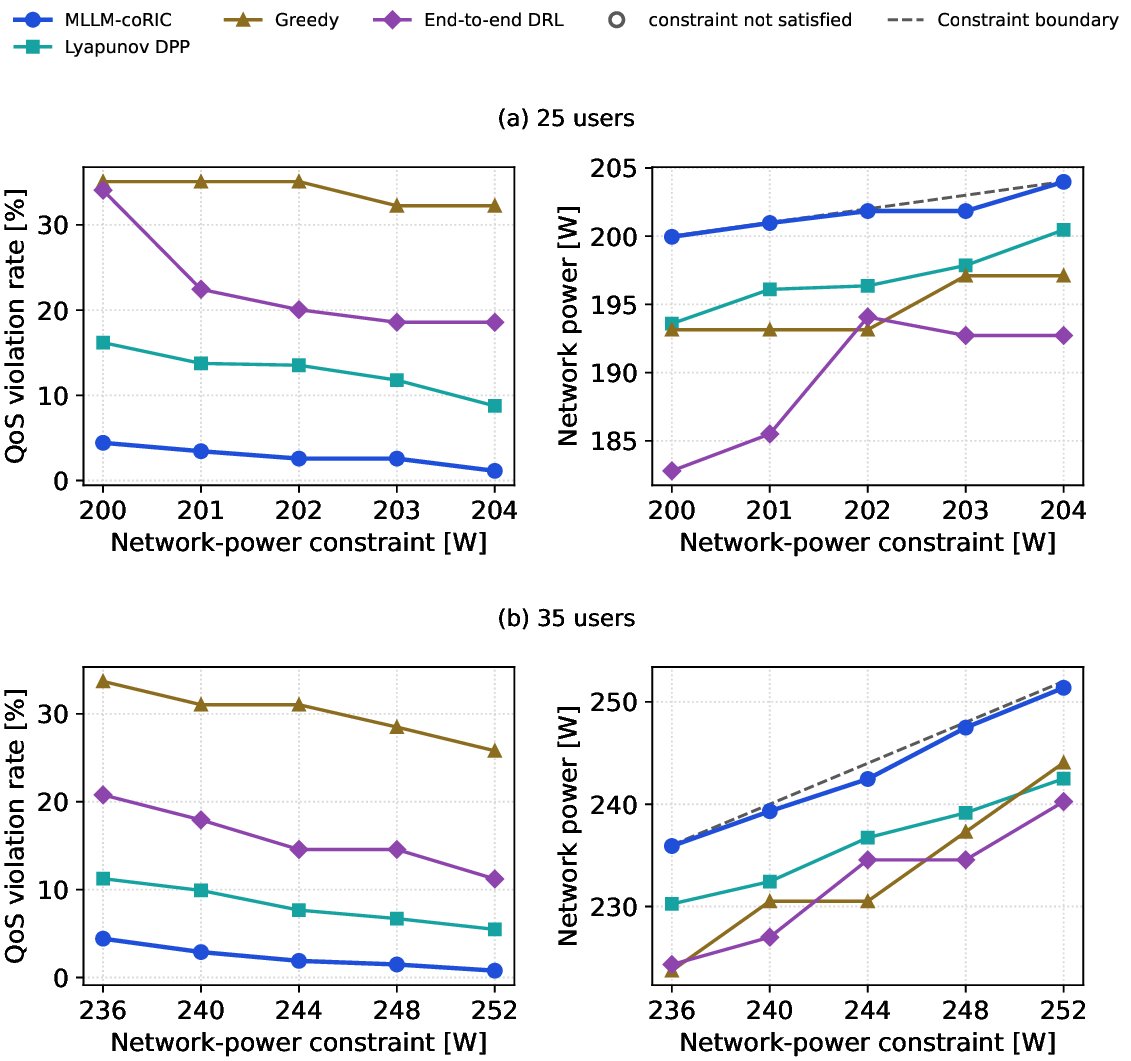}
\caption{QoS violation rate and network power under the network-power constraint $\overline P$ ($\Problem{3}$): (a) $25$ users, (b) $35$ users. Dashed line: the constraint boundary $P_{\rm tot}=\overline P$. Hollow markers miss the constraint.}
\label{fig:sw_pow}
\end{figure}

\begin{figure}[t]
\centering
\includegraphics[width=\columnwidth]{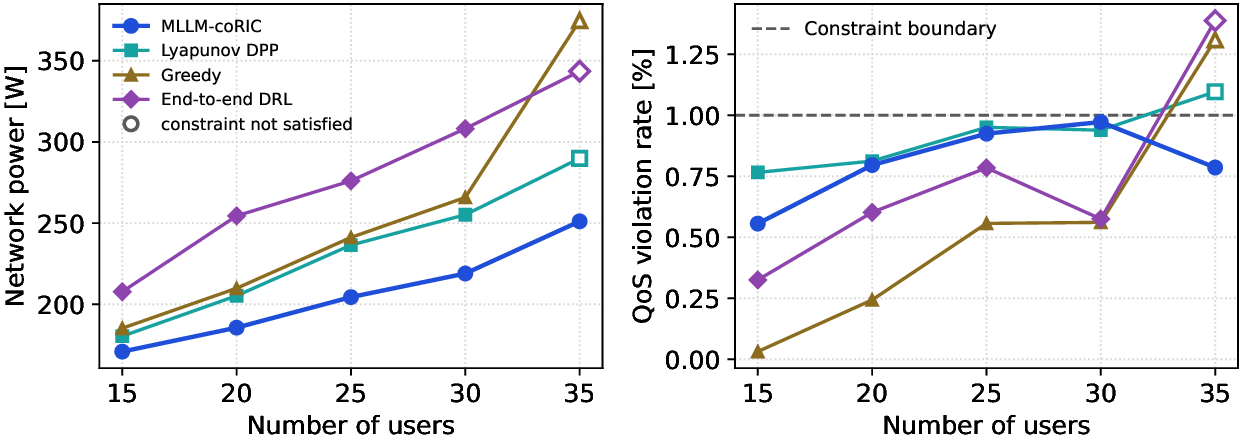}
\caption{Network power and QoS violation rate with respect to the number of users under $\Problem{2}$ with $\overline V=1\%$. Hollow markers miss the constraint.}
\label{fig:dens}
\end{figure}

\begin{figure*}[t]
\centering
\includegraphics[width=\textwidth]{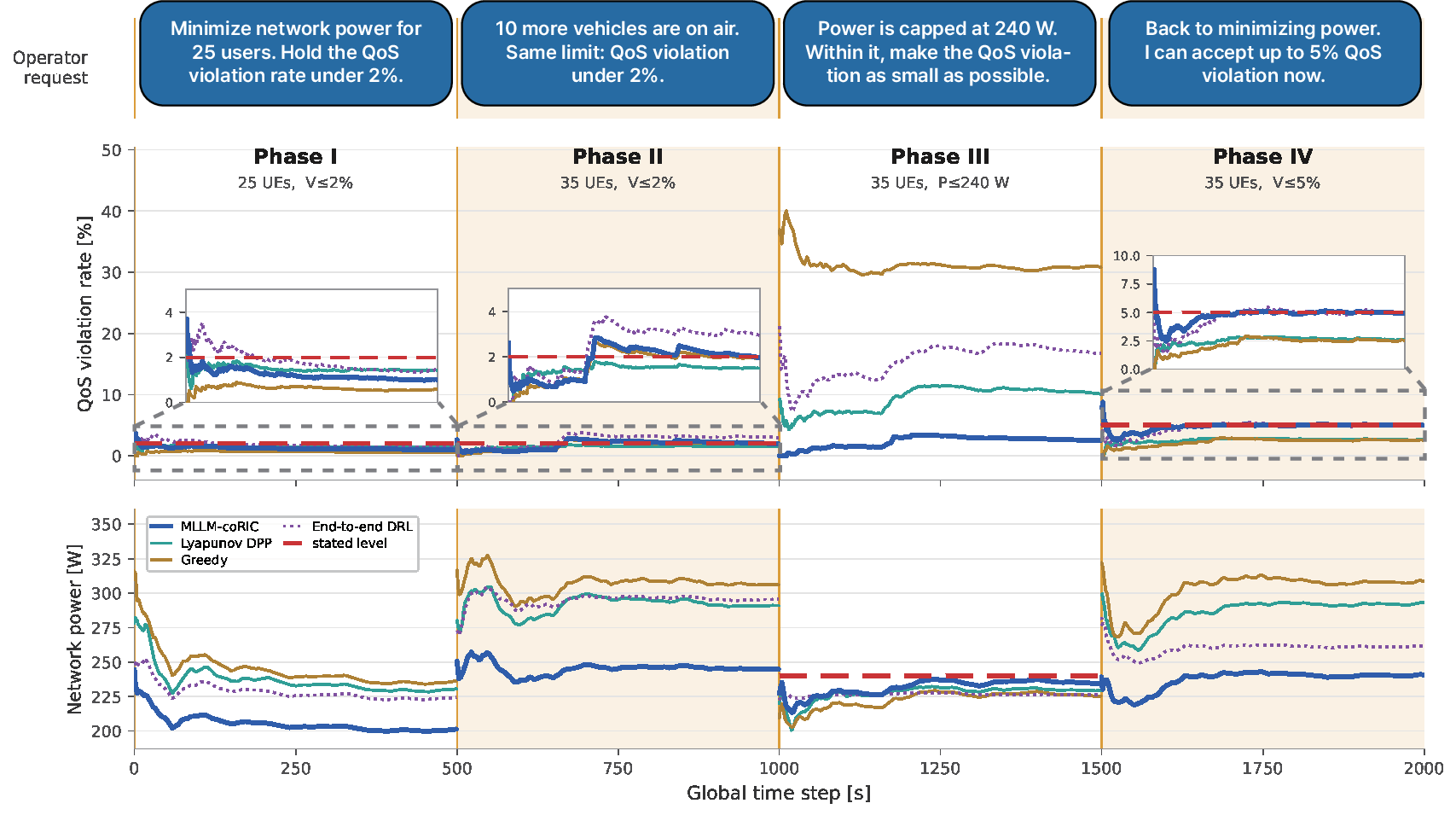}
\caption{One deployment across four phases on a global time step, with the operator's request for each phase above the axes. Curves are running averages within a phase; the red line is the stated level and steps with the request.}
\label{fig:phases}
\end{figure*}

The results for $\Problem{3}$ further demonstrate the problem-solving capability of MLLM-coRIC when $\mathrm{V}$ is minimized subject to the network-power constraint. As shown in Fig.~\ref{fig:sw_pow}, MLLM-coRIC consistently achieves the lowest QoS violation rate while operating close to the prescribed power limit. Although all methods are configured to solve the same constrained problem, the baselines generally yield operating points with larger power slack and correspondingly higher QoS violation rates. In contrast, MLLM-coRIC explicitly evaluates candidate resource--power pairs under the requirement-dependent loss, allowing additional power to be used whenever it further reduces $\mathrm{V}$ without violating the prescribed constraint. Thus, MLLM-coRIC makes more effective use of the feasible power budget, remaining within $1.5$\,W and $1.2$\,W of the prescribed limits at $35$ and $25$ users, respectively, while attaining the lowest QoS violation rate across the tested power budgets.

The simulation environment is fixed to $\Problem{2}$ with $\overline V=1\%$, while the number of users is changed to evaluate the problem-solving capability of MLLM-coRIC under various network loads. In Fig.~\ref{fig:dens}, MLLM-coRIC consistently achieves the lowest network power across the entire load range and remains feasible even under the most demanding scenario with $35$ users, whereas all three baselines violate the QoS constraint. As the user load increases, satisfying the same QoS constraint becomes more challenging because more radio resources and transmit power are required to serve the increased traffic. Nevertheless, MLLM-coRIC consistently finds feasible resource--power allocations while maintaining the lowest network power. In contrast, the baselines violate the QoS constraint under the challenging $35$-user load scenario, indicating that their solutions become less effective as the constrained optimization problem becomes more demanding.

\subsection{Adaptation to Dynamic Environment and Operator Requirements}
\label{sec:phases}

The next question is whether a single deployed controller can adapt when the network load and operator requirement change during operation. As shown in Fig.~\ref{fig:phases}, a single deployment is executed through four consecutive phases without policy retraining or manual retuning:
\begin{align*}
\text{Phase I:}\quad
& \min \; P_{\rm tot}
&& \text{s.t.}\quad \mathrm{V} \le 2\%, 
&& |\mathcal{U}_t| = 25,
\\
\text{Phase II:}\quad
& \min \; P_{\rm tot}
&& \text{s.t.}\quad \mathrm{V} \le 2\%, 
&& |\mathcal{U}_t| = 35,
\\
\text{Phase III:}\quad
& \min \; \mathrm{V}
&& \text{s.t.}\quad P_{\rm tot} \le 240~\mathrm{W}, 
&& |\mathcal{U}_t| = 35,
\\
\text{Phase IV:}\quad
& \min \; P_{\rm tot}
&& \text{s.t.}\quad \mathrm{V} \le 5\%, 
&& |\mathcal{U}_t| = 35.
\end{align*}
Phase~II therefore isolates adaptation to an increased network load under the same requirement, whereas Phase~III changes the optimization goal by exchanging the roles of the objective and constraint. Phase~IV restores power minimization while relaxing the QoS requirement. The trajectory, channel model, interference coupling, A1 cadence, and power model remain fixed throughout the sequence, so the observed changes are induced only by the network load and operator requirement. For the end-to-end DRL baseline, the model is retrained for each requirement to report its best achievable performance, although such retraining is impractical for online operation.

MLLM-coRIC adapts to changes in both the network load and the operator requirement across all four phases without retraining or manual retuning. From Phase~I to Phase~II, the increased user load requires higher network power, yet MLLM-coRIC continues to satisfy the QoS constraint on average while operating close to the stated boundary. Phase~III presents a substantially more challenging operating condition due to the stringent power constraint. Even under this limited power budget, MLLM-coRIC maintains a very low QoS violation rate, whereas all baselines suffer pronounced QoS degradation, with substantially higher violation rates. In Phase~IV, MLLM-coRIC again shifts its operating point toward the relaxed QoS boundary and reduces network power, while the baselines retain larger QoS margins at higher power consumption. It is noteworthy that the end-to-end DRL controller is retrained whenever the operator request changes, which is impractical for online RAN operation. Moreover, despite this retraining, the DRL controller still yields a worse operating point with respect to the stated objective than MLLM-coRIC. In contrast, MLLM-coRIC accommodates all phase transitions simply by updating the operator-request input to the MLLM agent, without retraining or manual retuning of the deployed controller.

\subsection{Forecast-Aided Proactive Provisioning}
\label{sec:proactive}

\begin{figure*}[t]
\centering
\includegraphics[width=\textwidth]{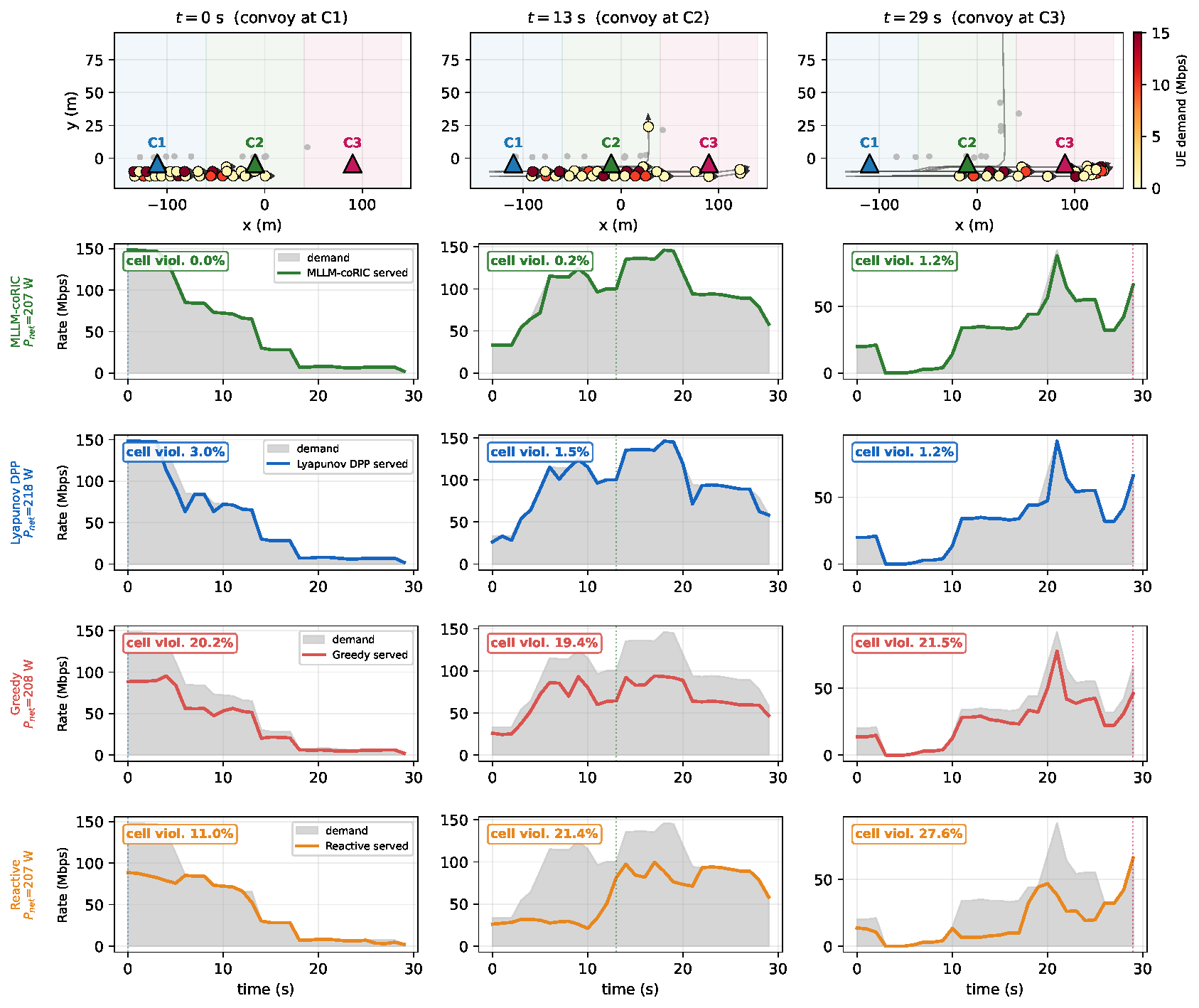}
\caption{Qualitative comparison in the CARLA Town05 traveling-hotspot scenario under matched network power. Top: snapshots as the vehicle convoy traverses the three cells, with vehicles colored by requested rate. Bottom: delivered and requested rates for each method over one representative $30$\,s episode.}
\label{fig:carla}
\end{figure*}

Fig.~\ref{fig:carla} further illustrates how prediction contributes to the resulting control performance. For a fair comparison, the power constraint of each baseline is configured such that its resulting network power is as close as possible to that of MLLM-coRIC. Under these comparable power-consumption levels, MLLM-coRIC tracks the requested traffic rates more closely as the users migrate across cells, resulting in substantially fewer QoS violations than the baselines. In particular, the reactive scheme begins provisioning resources only after the UE demand becomes locally observable. Because the PRB budget is subject to the ramp constraint, its resource allocation cannot increase quickly enough to follow a rapid rise in demand, which leads to temporary service deficits. In contrast, MLLM-coRIC exploits the predicted network evolution before the demand materializes. At the Non-RT RIC, the network-wide RF context, predicted future load, and operator requirement are jointly considered to synthesize a requirement-dependent loss from a holistic network perspective. The resulting future-aware criterion enables the Near-RT executor to provision resources in advance and coordinate transmit power as the hotspot moves across the network.

\subsection{Ablation Study}
\label{sec:ablation}

\begin{figure}[t]
\centering
\includegraphics[width=\columnwidth]{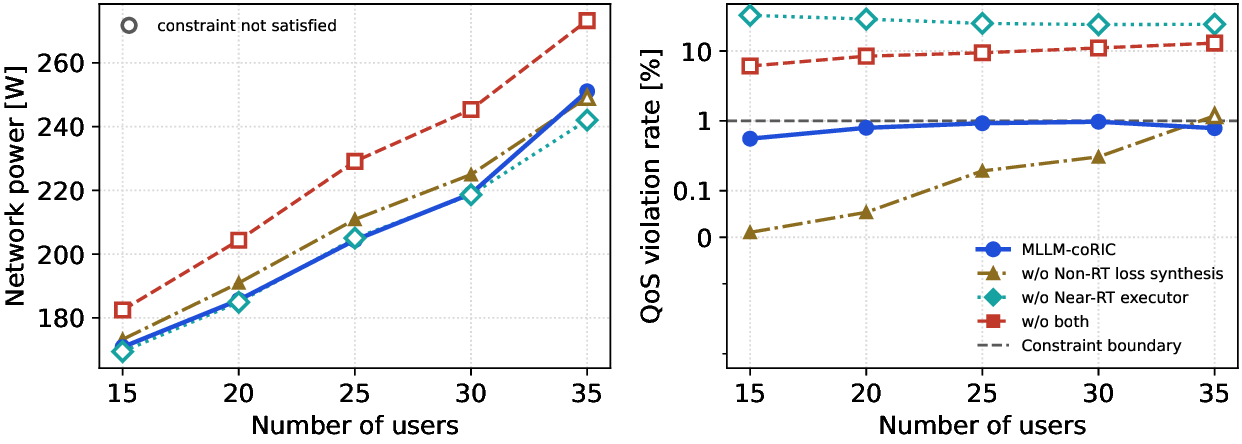}
\caption{Component ablation under $\Problem{2}$ at $\overline V=1\%$. Left: average network power consumption. Right: empirical QoS violation rate against the constraint boundary. Hollow markers indicate constraint violations.}
\label{fig:abl}
\end{figure}

The ablation in Fig.~\ref{fig:abl} clarifies the complementary roles of the Non-RT loss synthesis and the Near-RT executor under $\Problem{2}$ at $\overline V=1\%$. Without adaptive loss synthesis, fixed coefficients fail to maintain a consistent power--QoS tradeoff as the network load changes: they are overly conservative under lighter loads but become insufficient when the network is heavily loaded. In contrast, removing the Near-RT executor leads to severe QoS degradation even when a comparable amount of network power is consumed, showing that an appropriately synthesized loss alone does not guarantee effective radio control. MLLM-coRIC avoids both limitations by adapting the loss to the current requirement and network context at the Non-RT timescale and then translating that loss into state-dependent resource and power allocation at the Near-RT timescale. This result indicates that the performance gain of MLLM-coRIC arises from the coordinated operation of context-aware loss construction, feedback-driven refinement, and loss-conditioned radio control rather than from any single component in isolation. 

\begin{figure}[t]
\centering
\includegraphics[width=\columnwidth]{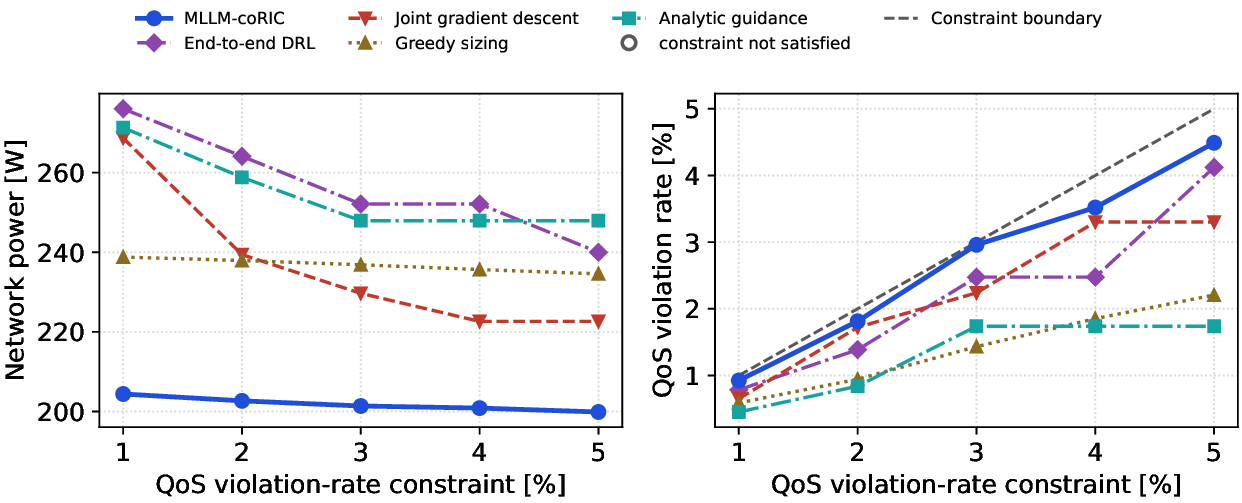}
\caption{Near-RT executor comparison under $\Problem{2}$ at $25$ users. Left: average network power consumption. Right: empirical QoS violation rate against the constraint boundary. Hollow markers indicate constraint violations.}
\label{fig:rahead}
\end{figure}

The Near-RT ablation study in Fig.~\ref{fig:rahead} compares MLLM-coRIC with a fully learning-based end-to-end DRL scheme and non-learning alternatives: joint gradient descent, which jointly optimizes the PRB budget and transmit power using the differentiable surrogate; greedy sizing; and analytic guidance, which applies forecast-guided PRB provisioning with full transmit power. MLLM-coRIC consistently operates near the requested QoS boundary while consuming the lowest network power across the constraint sweep. This result supports the proposed hybrid design: learning handles resource allocation, which must anticipate future load under the ramp constraint, whereas transmit power is directly optimized from the synthesized loss using the current channel and interference state. By assigning learning and model-based optimization to the decisions for which they are best suited, MLLM-coRIC achieves a more efficient power--QoS operating point than either approach alone.

\subsection{Computational Complexity}
\begin{figure}[t]
\centering
\includegraphics[width=\columnwidth]{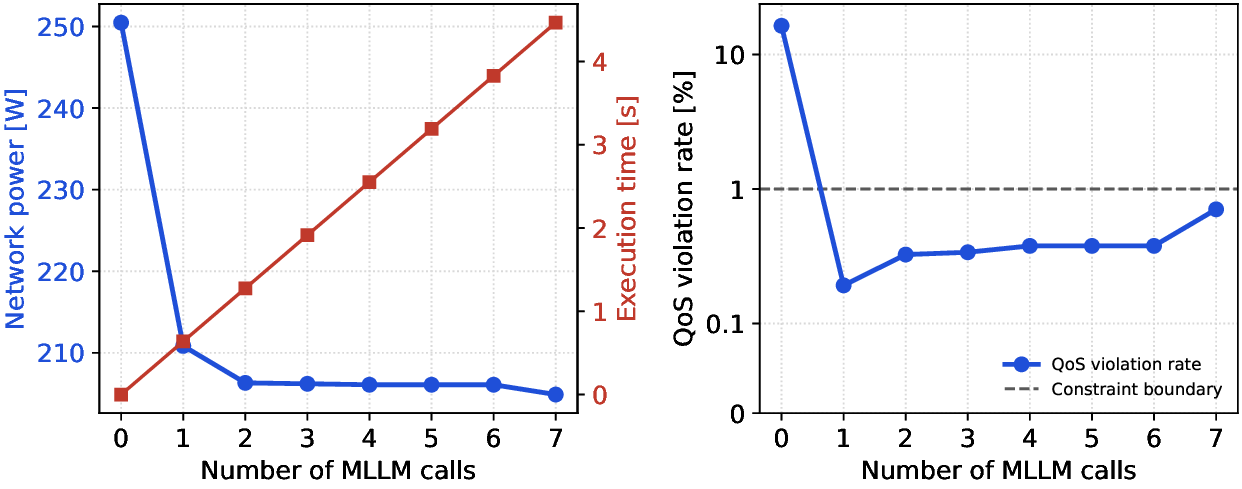}
\caption{MLLM-guided refinement for $\Problem{2}$ with $\overline V=1\%$. Left: network power and cumulative MLLM execution time versus the number of MLLM calls. Right: QoS violation rate against the constraint boundary.}
\label{fig:complexity}
\end{figure}

Fig.~\ref{fig:complexity} evaluates the computational overhead of the MLLM-guided refinement for $\Problem{2}$ with $\overline V=1\%$. With zero MLLM calls, the MLP initialization alone yields a highly infeasible operating point, whereas a single MLLM refinement reduces the QoS violation rate below the stated constraint while substantially lowering network power. Additional calls further improve the operating point, although the marginal gain becomes smaller as refinement proceeds. Meanwhile, the cumulative MLLM execution time increases approximately linearly and remains below $5$\,s even after seven calls. These results indicate that only a small number of MLLM refinement calls is sufficient to obtain a feasible and efficient control criterion with modest computational overhead.

\section{Conclusion}
This paper proposed MLLM-coRIC, a hierarchical O-RAN control framework that treats the operator-defined optimization problem as a runtime input rather than a fixed design-time specification. The key advantage of MLLM-coRIC lies in separating problem interpretation from fast radio execution. At the Non-RT RIC, the MLLM interprets the operator requirement together with the predicted network evolution and measured optimization outcomes to construct the numerical control loss, rather than directly generating radio actions. At the Near-RT RIC, the resulting loss is executed through a hybrid controller that uses learning for ramp-constrained resource provisioning and model-based optimization for interference-aware power control. This separation allows each component to address the decision structure for which it is best suited, while enabling different objectives, constraints, and operating conditions to be handled within a common control framework without redesigning a task-specific controller. The evaluation confirms that this design yields effective constrained solutions across different optimization problems and network loads, while requiring only a small number of MLLM-guided refinement rounds.

Future work will investigate the proposed framework in larger-scale and real O-RAN deployments, where imperfect measurements, forecasting errors, and interface delays may affect the loss-synthesis and control processes. It will also be of interest to extend the framework to a broader set of RAN control variables and operator-defined objectives, while further reducing the computational overhead of the Non-RT refinement procedure.

\bibliographystyle{IEEEtran}
\bibliography{bibtex}

@STRING{IEEE_J_TVT         = "{IEEE} Trans. Veh. Technol."}

@STRING{IEEE_J_TSP         = "{IEEE} Trans. Signal Process."}

@STRING{IEEE_J_CL       = "{IEEE} Commun. Lett."}

@STRING{IEEE_J_JSAC       = "{IEEE} J. Sel. Areas Commun."}

@STRING{IEEE_J_TCOM        = "{IEEE} Trans. Commun."}

@STRING{IEEE_J_TWC       = "{IEEE} Trans. Wireless Commun."}

@STRING{IEEE_S_COM        = "{IEEE} Commun. Surveys Tuts."}

@STRING{IEEE_J_TMC          = "{IEEE} Trans. Mobile Comput."}

@STRING{IEEE_J_PROC       = "Proc. {IEEE}"}

@STRING{IEEE_M_COM        = "{IEEE} Commun. Mag."}

@STRING{IEEE_M_VT         = "{IEEE} Veh. Technol. Mag."}

@STRING{IEEE_M_WC         = "{IEEE} Wireless Commun."}

@article{polese2023understanding,
  author  = {Polese, Michele and Bonati, Leonardo and D'Oro, Salvatore and
             Basagni, Stefano and Melodia, Tommaso},
  title   = {Understanding {O-RAN}: Architecture, Interfaces, Algorithms,
             Security, and Research Challenges},
  journal = IEEE_S_COM,
  volume  = {25},
  number  = {2},
  pages   = {1376--1411},
  year    = {2023},
}

@article{earth2011,
  author  = {Auer, G. and Giannini, V. and Desset, C. and G{\'o}dor, I. and
             Skillermark, P. and Olsson, M. and Imran, M. A. and Sabella, D.
             and Gonzalez, M. J. and Blume, O. and Fehske, A.},
  title   = {How Much Energy Is Needed to Run a Wireless Network?},
  journal = IEEE_M_WC,
  volume  = {18},
  number  = {5},
  pages   = {40--49},
  year    = {2011},
}

@techreport{ts22186,
  author      = {{3GPP}},
  title       = {Service Requirements for Enhanced {V2X} Scenarios},
  institution = {3rd Generation Partnership Project (3GPP)},
  number      = {TS 22.186, V15.4.0},
  year        = {2018},
}

@inproceedings{cho2014learning,
  title={Learning Phrase Representations using {RNN} Encoder--Decoder for Statistical Machine Translation},
  author={Cho, Kyunghyun and others},
  booktitle={Proceedings of the 2014 Conference on Empirical Methods in Natural Language Processing ({EMNLP})},
  pages={1724--1734},
  year={2014}
}

@ARTICLE{Letaief19_CM,
  author={Letaief, Khaled B. and Chen, Wei and Shi, Yuanming and Zhang, Jun and Zhang, Ying-Jun Angela},
  journal=IEEE_M_COM, 
  title={The Roadmap to {6G}: {AI} Empowered Wireless Networks}, 
  year={2019},
  volume={57},
  number={8},
  pages={84-90},
  doi={10.1109/MCOM.2019.1900271}}

@ARTICLE{Shi23_COMST,
  author={Shi, Yandong and others},
  journal=IEEE_S_COM, 
  title={Machine Learning for Large-Scale Optimization in {6G} Wireless Networks}, 
  year={2023},
  volume={25},
  number={4},
  pages={2088-2132},
  doi={10.1109/COMST.2023.3300664}}

@ARTICLE{Shafin20_WC,
  author={Shafin, Rubayet and others},
  journal=IEEE_M_WC, 
  title={Artificial Intelligence-Enabled Cellular Networks: A Critical Path to Beyond-{5G} and {6G}}, 
  year={2020},
  volume={27},
  number={2},
  pages={212-217},
  doi={10.1109/MWC.001.1900323}}

@ARTICLE{Yang25_VTM,
  author={Yang, Hyun Jong and Kim, Hyunsoo and Noh, Hyeonho and Kim, Seungnyun and Shim, Byonghyo},
  journal=IEEE_M_VT, 
  title={Large Multimodal Model-Empowered Task-Oriented Autonomous Communications: Design Methodology and Implementation Challenges}, 
  year={2025},
  volume={},
  number={},
  pages={2-13},
  doi={10.1109/MVT.2025.3626115}}

@ARTICLE{Polese23_COMST,
  author={Polese, Michele and Bonati, Leonardo and D\'oro, Salvatore and Basagni, Stefano and Melodia, Tommaso},
  journal=IEEE_S_COM, 
  title={Understanding {O-RAN}: Architecture, Interfaces, Algorithms, Security, and Research Challenges}, 
  year={2023},
  volume={25},
  number={2},
  pages={1376-1411},
  doi={10.1109/COMST.2023.3239220}}

@ARTICLE{Bonati21_COMM,
  author={Bonati, Leonardo and D\'oro, Salvatore and Polese, Michele and Basagni, Stefano and Melodia, Tommaso},
  journal=IEEE_M_COM, 
  title={Intelligence and Learning in {O-RAN} for Data-Driven NextG Cellular Networks}, 
  year={2021},
  volume={59},
  number={10},
  pages={21-27},
  doi={10.1109/MCOM.101.2001120}}

@ARTICLE{doro24_TMC,
  author={D\'oro, Salvatore and Bonati, Leonardo and Polese, Michele and Melodia, Tommaso},
  journal=IEEE_J_TMC, 
  title={{OrchestRAN}: Orchestrating Network Intelligence in the Open {RAN}}, 
  year={2024},
  volume={23},
  number={7},
  pages={7952-7968},
  doi={10.1109/TMC.2023.3342711}}

@ARTICLE{Tsampazi25_TMC,
  author={Tsampazi, Maria and others},
  journal=IEEE_J_TMC, 
  title={{PandORA}: Automated Design and Comprehensive Evaluation of Deep Reinforcement Learning Agents for Open {RAN}}, 
  year={2025},
  volume={24},
  number={4},
  pages={3223-3240},
  doi={10.1109/TMC.2024.3505781}}

@ARTICLE{Bao26_CM,
  author={Bao, Lingyan and Yun, Sinwoong and Lee, Jemin and Quek, Tony Q.S.},
  journal=IEEE_M_COM, 
  title={{LLM-hRIC}: {LLM}-Empowered Hierarchical {RAN} Intelligent Control for {O-RAN}}, 
  year={2026},
  volume={},
  number={},
  pages={1-7},
  doi={10.1109/MCOM.001.2500315}}

@ARTICLE{Zappone19_TCOM,
  author={Zappone, Alessio and Di Renzo, Marco and Debbah, M\'erouane},
  journal=IEEE_J_TCOM, 
  title={Wireless Networks Design in the Era of Deep Learning: Model-Based, {AI}-Based, or Both?}, 
  year={2019},
  volume={67},
  number={10},
  pages={7331-7376},
  doi={10.1109/TCOMM.2019.2924010}}

@ARTICLE{Dai25_TMC,
  author={Dai, Jiongyu and others},
  journal=IEEE_J_TMC, 
  title={{O-RAN-Enabled} Intelligent Network Slicing to Meet Service-Level Agreement ({SLA})}, 
  year={2025},
  volume={24},
  number={2},
  pages={890-906},
  doi={10.1109/TMC.2024.3476338}}

@ARTICLE{Wu21_JSAC,
  author={Wu, Wen and others},
  journal=IEEE_J_JSAC, 
  title={Dynamic {RAN} Slicing for Service-Oriented Vehicular Networks via Constrained Learning}, 
  year={2021},
  volume={39},
  number={7},
  pages={2076-2089},
  doi={10.1109/JSAC.2020.3041405}}

@ARTICLE{Abedin22_TVT,
  author={Abedin, Sarder Fakhrul and Mahmood, Aamir and Tran, Nguyen H. and Han, Zhu and Gidlund, Mikael},
  journal=IEEE_J_TVT, 
  title={Elastic {O-RAN} Slicing for Industrial Monitoring and Control: A Distributed Matching Game and Deep Reinforcement Learning Approach}, 
  year={2022},
  volume={71},
  number={10},
  pages={10808-10822},
  doi={10.1109/TVT.2022.3188217}}

@ARTICLE{Ju22_TWC,
  author={Ju, Hyungyu and Kim, Seungnyun and Kim, Youngjoon and Shim, Byonghyo},
  journal=IEEE_J_TWC, 
  title={Energy-Efficient Ultra-Dense Network With Deep Reinforcement Learning}, 
  year={2022},
  volume={21},
  number={8},
  pages={6539-6552},
  doi={10.1109/TWC.2022.3150425}}

@ARTICLE{Noh24_CL,
  author={Noh, Hyeonho and Lee, Harim and Yang, Hyun Jong},
  journal=IEEE_J_CL, 
  title={Joint Optimization on Uplink {OFDMA} and {MU-MIMO} for IEEE 802.11ax: Deep Hierarchical Reinforcement Learning Approach}, 
  year={2024},
  volume={28},
  number={8},
  pages={1800-1804},
  doi={10.1109/LCOMM.2024.3402959}}

@ARTICLE{Shen21_JSAC,
  author={Shen, Yifei and Shi, Yuanming and Zhang, Jun and Letaief, Khaled B.},
  journal=IEEE_J_JSAC, 
  title={Graph Neural Networks for Scalable Radio Resource Management: Architecture Design and Theoretical Analysis}, 
  year={2021},
  volume={39},
  number={1},
  pages={101-115},
  doi={10.1109/JSAC.2020.3036965}}

@ARTICLE{Noh25_TVT,
  author={Noh, Hyeonho and Shim, Byonghyo and Yang, Hyun Jong},
  journal=IEEE_J_TVT, 
  title={Adaptive Resource Allocation Optimization Using Large Language Models in Dynamic Wireless Environments}, 
  year={2025},
  volume={74},
  number={10},
  pages={16630-16635},
  doi={10.1109/TVT.2025.3572440}}

@ARTICLE{Wu22_COMST,
  author={Wu, Yulei and Dai, Hong-Ning and Wang, Haozhe and Xiong, Zehui and Guo, Song},
  journal=IEEE_S_COM, 
  title={A Survey of Intelligent Network Slicing Management for Industrial {IoT}: Integrated Approaches for Smart Transportation, Smart Energy, and Smart Factory}, 
  year={2022},
  volume={24},
  number={2},
  pages={1175-1211},
  doi={10.1109/COMST.2022.3158270}}

@ARTICLE{Polese24_JSAC,
  author={Polese, Michele and others},
  journal=IEEE_J_JSAC,
  title={Empowering the {6G} Cellular Architecture With Open {RAN}},
  year={2024},
  volume={42},
  number={2},
  pages={245-262},
  doi={10.1109/JSAC.2023.3334610}}

@ARTICLE{Polese23_TMC,
  author={Polese, Michele and Bonati, Leonardo and D'Oro, Salvatore and Basagni, Stefano and Melodia, Tommaso},
  journal=IEEE_J_TMC,
  title={{ColO-RAN}: Developing Machine Learning-Based {xApps} for Open {RAN} Closed-Loop Control on Programmable Experimental Platforms},
  year={2023},
  volume={22},
  number={10},
  pages={5787-5800},
  doi={10.1109/TMC.2022.3188885}}

@ARTICLE{Lacava24_TMC,
  author={Lacava, Andrea and others},
  journal=IEEE_J_TMC,
  title={Programmable and Customized Intelligence for Traffic Steering in {5G} Networks Using Open {RAN} Architectures},
  year={2024},
  volume={23},
  number={4},
  pages={2882-2897},
  doi={10.1109/TMC.2023.3266642}}

@ARTICLE{Apostolakis24_JSAC,
  author={Apostolakis, Nikolaos and others},
  journal=IEEE_J_JSAC,
  title={{ATHENA}: Machine Learning and Reasoning for Radio Resources Scheduling in {vRAN} Systems},
  year={2024},
  volume={42},
  number={2},
  pages={263-279},
  doi={10.1109/JSAC.2023.3336175}}

@ARTICLE{Eisen20_TSP,
  author={Eisen, Mark and Ribeiro, Alejandro},
  journal=IEEE_J_TSP,
  title={Optimal Wireless Resource Allocation With Random Edge Graph Neural Networks},
  year={2020},
  volume={68},
  pages={2977-2991},
  doi={10.1109/TSP.2020.2988255}}

@ARTICLE{Chowdhury21_TWC,
  author={Chowdhury, Arindam and Verma, Gunjan and Rao, Chirag and Swami, Ananthram and Segarra, Santiago},
  journal=IEEE_J_TWC,
  title={Unfolding {WMMSE} Using Graph Neural Networks for Efficient Power Allocation},
  year={2021},
  volume={20},
  number={9},
  pages={6004-6017},
  doi={10.1109/TWC.2021.3071480}}

@ARTICLE{Naderializadeh21_TWC,
  author={Naderializadeh, Navid and Sydir, Jaroslaw J. and Simsek, Meryem and Nikopour, Hosein},
  journal=IEEE_J_TWC,
  title={Resource Management in Wireless Networks via Multi-Agent Deep Reinforcement Learning},
  year={2021},
  volume={20},
  number={6},
  pages={3507-3523},
  doi={10.1109/TWC.2021.3051163}}

@ARTICLE{Mei21_TCOM,
  author={Mei, Jie and others},
  journal=IEEE_J_TCOM,
  title={Intelligent Radio Access Network Slicing for Service Provisioning in {6G}: A Hierarchical Deep Reinforcement Learning Approach},
  year={2021},
  volume={69},
  number={9},
  pages={6063-6078},
  doi={10.1109/TCOMM.2021.3090423}}

@ARTICLE{AyalaRomero22_TMC,
  author={Ayala-Romero, Jose A. and others},
  journal=IEEE_J_TMC, 
  title={{vrAIn}: Deep Learning Based Orchestration for Computing and Radio Resources in {vRANs}}, 
  year={2022},
  volume={21},
  number={7},
  pages={2652-2670},
  doi={10.1109/TMC.2020.3043100}}

@ARTICLE{Naderializadeh23_TSP,
  author={NaderiAlizadeh, Navid and Eisen, Mark and Ribeiro, Alejandro},
  journal=IEEE_J_TSP,
  title={Learning Resilient Radio Resource Management Policies With Graph Neural Networks},
  year={2023},
  volume={71},
  pages={995-1009},
  doi={10.1109/TSP.2023.3255543}}

@ARTICLE{Nguyen22_PROC,
  author={Nguyen, Cong T. and others},
  journal=IEEE_J_PROC,
  title={Transfer Learning for Wireless Networks: A Comprehensive Survey},
  year={2022},
  volume={110},
  number={8},
  pages={1073-1115},
  doi={10.1109/JPROC.2022.3175942}}

@ARTICLE{Zhou25_COMST,
  author={Zhou, Hao and others},
  journal=IEEE_S_COM,
  title={Large Language Model ({LLM}) for Telecommunications: A Comprehensive Survey on Principles, Key Techniques, and Opportunities},
  year={2025},
  volume={27},
  number={3},
  pages={1955-2005},
  doi={10.1109/COMST.2024.3465447}}

@ARTICLE{Maatouk24_CM,
  author={Maatouk, Ali and Piovesan, Nicola and Ayed, Fadhel and De Domenico, Antonio and Debbah, M{\'e}rouane},
  journal=IEEE_M_COM, 
  title={Large Language Models for Telecom: Forthcoming Impact on the Industry}, 
  year={2025},
  volume={63},
  number={1},
  pages={62-68},
  doi={10.1109/MCOM.001.2300473}}

@ARTICLE{Chen24_WC,
  author={Chen, Zirui and Zhang, Zhaoyang and Yang, Zhaohui},
  journal=IEEE_M_WC,
  title={Big {AI} Models for {6G} Wireless Networks: Opportunities, Challenges, and Research Directions},
  year={2024},
  volume={31},
  number={5},
  pages={164-172},
  doi={10.1109/MWC.014.2300404}}

@ARTICLE{Zhou25_WC,
  author={Zhou, Hao and others},
  journal=IEEE_M_WC, 
  title={Large Language Models for Wireless Networks: An Overview from the Prompt Engineering Perspective}, 
  year={2025},
  volume={32},
  number={4},
  pages={98-106},
  doi={10.1109/MWC.001.2400384}}

@ARTICLE{Leivadeas23_COMST,
  author={Leivadeas, Aris and Falkner, Matthias},
  journal=IEEE_S_COM, 
  title={A Survey on Intent-Based Networking}, 
  year={2023},
  volume={25},
  number={1},
  pages={625-655},
  doi={10.1109/COMST.2022.3215919}}

@ARTICLE{Xu24_WC,
  author={Xu, Minrui and others},
  journal=IEEE_M_WC, 
  title={When Large Language Model Agents Meet 6G Networks: Perception, Grounding, and Alignment}, 
  year={2024},
  volume={31},
  number={6},
  pages={63-71},
  doi={10.1109/MWC.005.2400019}}

@ARTICLE{Jiang24_WC,
  author={Jiang, Feibo and others},
  journal=IEEE_M_WC, 
  title={Large Language Model Enhanced Multi-Agent Systems for 6G Communications}, 
  year={2024},
  volume={31},
  number={6},
  pages={48-55},
  doi={10.1109/MWC.016.2300600}}

@ARTICLE{Park26_TVT,
  author={Park, Sojeong and Noh, Hyeonho and Yang, Hyun Jong},
  journal=IEEE_J_TVT, 
  title={Robust Transmission of Punctured Text With Large Language Model-Based Recovery}, 
  year={2026},
  volume={75},
  number={1},
  pages={1737-1742},
  doi={10.1109/TVT.2025.3595593}}

\end{document}